%% file: channel_dispersion_pdf-2.3.tex
\documentclass[tightenlines,twocolumn,preprintnumbers,amsmath,amssymb,showkeys,twoside,aps,floatfix,prb]{revtex4}
\usepackage{graphicx}% Include figure files
\usepackage{dcolumn}% Align table columns on decimal point
\usepackage{bm}% bold math
\usepackage{color}% for including pstex figures from Xfig
\usepackage[colorlinks,letterpaper,dvips]{hyperref}

\hypersetup{citecolor=blue,
            linkcolor=blue,
	    urlcolor=blue,
	    pdftitle=Probability density function modeling of scalar mixing
	             from concentrated sources in turbulent channel flow,
	    pdfauthor=Jozsef Bakosi,
	    pdfkeywords=turbulent channel flow; scalar dispersion;
	                probability density function method; particle method;
			Langevin equation; Monte-Carlo method}
\usepackage{hypernat}
\begin{document}

\newcommand{\bv}[1]{\mbox{\boldmath$#1$}}
\newcommand{\mean}[1]{\left<#1\right>}
\newcommand{\sd}[1]{\frac{\mathrm{D}#1}{\mathrm{D}t}}
\newcommand{\msd}[1]{\frac{\mathrm{\overline{D}}#1}{\mathrm{\overline{D}}t}}
\newcommand*{\Eqr}[1]{(\ref{#1})}
\newcommand*{\Eqre}[1]{Eq.~(\ref{#1})}
\newcommand*{\Eqrs}[1]{(\ref{#1})}
\newcommand*{\Eqres}[1]{Eqs.~(\ref{#1})}
\newcommand*{\Fig}[1]{fig.~\ref{#1}}
\newcommand*{\Figs}[1]{figs.~\ref{#1}}
\newcommand*{\Fige}[1]{Fig.~\ref{#1}}
\newcommand*{\Figse}[1]{Figs.~\ref{#1}}

\title{Probability density function modeling of scalar mixing\\from concentrated sources in turbulent channel flow}

\author{J. Bakosi}
\email{jbakosi@gmu.edu}
\author{P. Franzese}
\author{Z. Boybeyi}
\affiliation{College of Science, George Mason University, Fairfax, VA, 22030, USA}

\date{Sep.\ 29, 2007}


\begin{abstract}Dispersion of a passive scalar from concentrated sources in fully developed turbulent channel flow is studied with the probability density function (PDF) method. The joint PDF of velocity, turbulent frequency and scalar concentration is represented by a large number of Lagrangian particles. A stochastic near-wall PDF model combines the generalized Langevin model of \citet{Haworth_86} with Durbin's \citep{Durbin_93} method of elliptic relaxation to provide a mathematically exact treatment of convective and viscous transport with a non-local representation of the near-wall Reynolds stress anisotropy. The presence of walls is incorporated through the imposition of no-slip and impermeability conditions on particles without the use of damping or wall-functions. Information on the turbulent timescale is supplied by the gamma-distribution model of \citet{VanSlooten_98}. Two different micromixing models are compared that incorporate the effect of small scale mixing on the transported scalar: the widely used interaction by exchange with the mean (IEM) and the interaction by exchange with the conditional mean (IECM) model. Single-point velocity and concentration statistics are compared to direct numerical simulation and experimental data at \(\textit{Re}_\tau=1080\) based on the friction velocity and the channel half width. The joint model accurately reproduces a wide variety of conditional and unconditional statistics in both physical and composition space.
\end{abstract}

\maketitle

\section{Introduction}
In engineering industry and atmospheric transport and dispersion modeling there is an increasing use of computational methods to calculate complex turbulent flow fields. Many of these computations depend on the \(k\)--\(\varepsilon\) turbulence model \citep{Jones_72,Bacon_00}, while some are based on second-moment closures \citep{Rotta_51,Launder_75,Hanjalic_72,Speziale_91}. The aim of these statistical methods is to predict the first and second moments of the turbulent velocity field, respectively. In large-eddy simulation (LES) the large-scale three-dimensional unsteady motions are represented exactly, while the small-scale motions are parameterized. As long as the transport-controlling processes of interest (eg.\ mass, momentum and heat transfer in shear flows) are resolved, LES predictions can be expected to be insensitive to the details of residual-scale modeling. In applications such as high-Reynolds-number turbulent combustion or near-wall flows, however, where the important rate-controlling processes occur below the resolved scales, the residual-scale models directly influence the model predictions. Since there is no universally `best' methodology that is applicable for every type of practical flow, it is valuable to develop improvements for the full range of turbulence modeling  approaches.

The development of probability density function (PDF) methods is an effort to provide a higher-level statistical description of turbulent flows. The mean velocity and Reynolds stresses are statistics of (and can be obtained from) the PDF of velocity. In PDF methods, a transport equation is solved directly for the PDF of the turbulent velocity field, rather than for its moments as in Reynolds stress closures. Therefore, in principle, a more complete statistical description can be obtained. While for some flows (e.g.\ homogeneous turbulence) this higher-level description may provide little benefit over second moment closures, in general the fuller description is beneficial in allowing more processes to be treated exactly and in providing more information that can be used in the construction of closure models. Convection, for example, can be exactly represented mathematically in the PDF framework, eliminating the need for a closure assumption \citep{Pope_00}. Similarly, defining the joint PDF of velocity and species concentrations in a chemically reactive turbulent flow allows for the treatment of chemical reactions without the burden of closure assumptions for the highly nonlinear chemical source terms \citep{Fox_03}. This latter advantage has been one of the most important incentives for the development of PDF methods, since previous attempts to provide moment closures for this term resulted in errors of several orders of magnitude \citep{Pope_90}.

In the case of turbulent flows around complex geometries the presence of walls requires special treatment, since traditional turbulence models are developed for high Reynolds numbers and need to be modified in the vicinity of walls. This is necessary because the Reynolds number approaches zero at the wall, the highest shear rate occurs near the wall and the impermeability condition on the wall-normal velocity affects the flow up to an integral scale from the wall \citep{Hunt_78}. Possible modifications involve damping functions \citep{VanDriest_56,Lai_90,Craft_96,Rodi_93} or wall-functions \citep{Launder_74,Singhal_81,Rodi_80,Spalding_77}. In those turbulent flows where a higher level of statistical description is necessary close to walls, adequate representation of the near-wall anisotropy and inhomogeneity is crucial. \citet{Durbin_93} proposed a Reynolds stress closure to address these issues. In his model, the all-important process of pressure redistribution is modeled through an elliptic equation by analogy with the Poisson equation, which governs the pressure in incompressible flows. This represents the non-local effect of the wall on the Reynolds stresses through the fluctuating pressure terms. In an effort to extend PDF methods to wall-bounded turbulent flows, Durbin's elliptic relaxation method has been combined with the generalized Langevin model \citep{Haworth_86} by \citet{Dreeben_97,Dreeben_98}. With minor simplifications this approach is closely followed throughout the present study to model the joint PDF of the turbulent velocity field.

The dispersion of scalars (e.g.\ temperature, mass, etc.) in turbulent flows is relevant to a number of scientific phenomena including engineering combustion and atmospheric dispersion of pollutants. Reviews on the subject have been compiled by \citet{Shraiman_00} and \citet{Warhaft_00}. Several experimental studies have been carried out in order to better understand the behavior of transported scalars in homogeneous isotropic turbulence \citep{Warhaft_84,Stapountzis_86,Sawford_95}. A literature review of dispersion from a concentrated source in homogeneous but anisotropic turbulent shear flows is given by \citet{Karnik_89}. Inhomogeneous turbulence (e.g.\ the atmospheric boundary layer or any practical turbulent flow) adds a significant level of complexity to these cases. Extensive measurements of the mean, variance, intermittency, probability density functions and spectra of scalar have been made by \citet{Fackrell_82} in a turbulent boundary layer. One point statistics in turbulent channel flow have recently been reported by \citet{Lavertu_05}, whose experimental data are used as the main reference point throughout the current numerical modeling study.

Direct numerical simulation (DNS) has served as an important counterpart to measurements of turbulence at moderate Reynolds numbers, shedding light on quantities that are difficult to measure (e.g.\ Lagrangian statistics) and at locations where nearly impossible to measure (e.g.\ close to walls). Turbulent velocity statistics extracted from DNS of channel flow have been reported by \citet{Moser_99} and \citet{Abe_04}, while \citet{Vrieling_03} performed a DNS study of dispersion of plumes from single and double line sources.

A widely used model to incorporate the effects of small scale mixing on the scalar in the PDF framework is the interaction by exchange with the mean (IEM) model \citep{Villermaux_Devillon_72,Dopazo_OBrien_74}. While this model has the virtue of being simple and efficient, it fails to comply with several physical constraints and desirable properties of an ideal mixing model \citep{Fox_03}. Although a variety of other mixing models have been proposed to satisfy these properties \citep{Dopazo_94}, the IEM model remains widely used in practice. Recently, increased attention has been devoted to the interaction by exchange with the conditional mean (IECM) model. \citet{Sawford_04} has done a comparative study of scalar mixing from line sources in homogeneous turbulence employing both the IEM and IECM models, wherein he demonstrated that the largest differences between the two models occur in the near-field. He also investigated the two models in a double scalar mixing layer \citep{Sawford_06} with an emphasis on those conditional statistics that frequently require closure assumptions. Based on the IECM model, PDF micromixing models have been developed for dispersion of passive pollutants in the atmosphere by \citet{Luhar_et_al_05} and Cassiani et al.\ \cite{Cassiani_et_al_05,Cassiani_et_al_05b,Cassiani_07}. These authors compute scalar statistics in homogeneous turbulence and in neutral, convective and canopy boundary layer by assuming a joint PDF for the turbulent velocity field. However, no previous studies have been conducted on modeling the joint PDF of velocity and a passive scalar from a concentrated source in inhomogeneous flows.

We have developed a complete PDF-IECM model for a fully developed, turbulent, long-aspect-ratio channel flow, where a passive scalar is continuously released from concentrated sources. The joint PDF of velocity, characteristic turbulent frequency and concentration of a passive scalar is computed using stochastic equations. The flow is explicitly modeled down to the viscous sublayer by imposing only the no-slip and impermeability condition on particles without the use of damping, or wall-functions. The high-level inhomogeneity and anisotropy of the Reynolds stress tensor at the wall are captured by the elliptic relaxation method. A passive scalar is released from a concentrated source at the channel centerline and in the viscous wall-region. The effect of small-scale mixing on the scalar is mainly modeled by the IECM model. The performance and accuracy of the IECM model compared to the simpler, but more widely used IEM model are evaluated. Several one-point unconditional and conditional statistics are presented in both physical and composition spaces. An emphasis is placed on common approximations of those conditional statistics that require closure assumptions in concentration-only PDF methods, i.e.\ in methods that assume the underlying turbulent velocity field. The results are compared to the DNS data of \citet{Abe_04} and the experimental data of \citet{Lavertu_05}. The experiments were performed at two different Reynolds numbers (\(\textit{Re}_\tau\equiv u_\tau h/\nu=520\) and \(1080\) based on the friction velocity \(u_\tau\), the channel half width \(h\), and the kinematic viscosity \(\nu\)) in a high-aspect-ratio turbulent channel flow, measuring one point statistics of a scalar (temperature) emitted continuously at three different wall-normal source locations from concentrated line sources. Measurements were performed at six different downstream locations between \(4.0\leqslant x/h\leqslant22.0\).

The remainder of the paper is organized as follows. In Section \ref{sec:governing_equations}, the turbulence and micromixing models are described. A brief account of the underlying numerical methods, various implementation details together with wall-boundary conditions are presented in Section \ref{sec:num_method}. In Section \ref{sec:results}, one-point velocity statistics are compared to direct numerical simulation data at \(\textit{Re}_\tau=1080\), and a comparative assessment of the two micromixing models with analytical and experimental data is also given. Detailed statistics of scalar concentration calculated with the IECM micromixing model are presented. Finally, conclusions are summarized in Section \ref{sec:conclusion}.

\section{Governing equations}
\label{sec:governing_equations}
The governing equation for the motion of a viscous, incompressible fluid is the Navier-Stokes equation
\begin{equation}
\frac{\partial U_i}{\partial t} + U_j\frac{\partial U_i}{\partial x_j} + \frac{1}{\rho}\frac{\partial P}{\partial x_i} =  \nu\nabla^2U_i \label{eq:NavierStokes},
\end{equation}
where \(U_i\), \(P\), \(\rho\) and \(\nu\) are the Eulerian velocity, pressure, constant density and kinematic viscosity, respectively. From \Eqre{eq:NavierStokes} an exact transport equation can be derived for the one-point, one-time Eulerian joint PDF of velocity \(f(\bv{V};\bv{x},t)\) \citep{Pope_85,Pope_00},
\begin{equation}
\begin{split}
\frac{\partial f}{\partial t} + V_i\frac{\partial f}{\partial x_i} &= \nu\frac{\partial^2f}{\partial x_i\partial x_i} + \frac{1}{\rho}\frac{\partial\mean{P}}{\partial x_i}\frac{\partial f}{\partial V_i}\\
&\quad- \frac{\partial^2}{\partial V_i\partial V_j}\left[f\mean{\nu\frac{\partial U_i}{\partial x_k}\frac{\partial U_j}{\partial x_k}\Bigg|\bv{U}(\bv{x},t)=\bv{V}}\right]\\
&\quad+ \frac{\partial}{\partial V_i}\left[f\mean{\frac{1}{\rho}\frac{\partial p}{\partial x_i}\Bigg|\bv{U}(\bv{x},t)=\bv{V}}\right],
\end{split}
\label{eq:EulerianPDFforviscous}
\end{equation}
where \(\bv{V}\) is the sample space variable of the stochastic velocity field \(\bv{U}(\bv{x},t)\) and the pressure \(P\) has been decomposed into mean \(\mean{P}\) and fluctuating part \(p\). Since the PDF \(f(\bv{V};\bv{x},t)\) contains all one-point statistics, mean velocity and Reynolds stresses are readily available through
\begin{eqnarray}
\mean{U_i}&=&\int V_i f\mathrm{d}\bv{V},\\
\mean{u_iu_j}&=&\int (V_i-\mean{U_i})(V_j-\mean{U_j}) f\mathrm{d}\bv{V},
\end{eqnarray}
where the fluctuation \(u_i=V_i-\mean{U_i}\) and the integrals are taken over all the three-dimensional sample space of the velocity field. While in principle, after modeling the unclosed conditional statistics, \Eqre{eq:EulerianPDFforviscous} could be solved with traditional numerical techniques, such as the finite difference or finite element method, the high dimensionality of the equation prevents an efficient solution with these methods. In general, Monte-Carlo techniques are computationally more efficient for problems with such high dimensions. In PDF methods, therefore, a Lagrangian description has been preferred. The Navier-Stokes equation \eqref{eq:NavierStokes} is widely used to model incompressible flows in the Eulerian framework. An equivalent model in the Lagrangian framework can be written as a system of governing equations for Lagrangian particle locations \(\mathcal{X}_i\) and velocities \(\mathcal{U}_i\) \citep{Dreeben_97}
\begin{eqnarray}
\mathrm{d}\mathcal{X}_i&=&\mathcal{U}_i\mathrm{d}t + \left(2\nu\right)^{1/2}\mathrm{d}W_i\label{eq:Lagrangian-position-exact}\\
\mathrm{d}\mathcal{U}_i(t)&=&-\frac{1}{\rho}\frac{\partial P}{\partial x_i}\mathrm{d}t + 2\nu\frac{\partial^2U_i}{\partial x_j\partial x_j}\mathrm{d}t+\left(2\nu\right)^{1/2}\frac{\partial U_i}{\partial x_j}\mathrm{d}W_j,\label{eq:Lagrangian-velocity-exact}
\end{eqnarray}
where the isotropic Wiener process \citep{Gardiner_04} \(\mathrm{d}W_i\), which is a known stochastic process with zero mean and variance \(\mathrm{d}t\), is identical in both equations (the same exact series of Gaussian random numbers) and it is understood that the Eulerian fields on the right hand side are evaluated at the particle locations \(\mathcal{X}_i\). Equation (\ref{eq:Lagrangian-position-exact}) governs the position of a stochastic fluid particle which undergoes both convective and molecular motion. In other words, besides convection the particle diffuses in physical space with coefficient \(\nu\), thus it carries momentum as molecules do with identical statistics, as in Brownian motion \citep{Einstein_26}. Each of the above three descriptions, i.e.\ \Eqres{eq:NavierStokes}, \Eqr{eq:EulerianPDFforviscous} and (\ref{eq:Lagrangian-position-exact}-\ref{eq:Lagrangian-velocity-exact}), represents the viscous stress exactly. A remarkable feature of the PDF formulation is that the effects of convection and viscous diffusion, fundamental processes in near-wall turbulent flows, have exact mathematical representations, therefore need no closure assumptions. If Reynolds decomposition is applied to \Eqre{eq:Lagrangian-velocity-exact}
\begin{equation}
\begin{split}
\mathrm{d}\mathcal{U}_i(t)&=-\frac{1}{\rho}\frac{\partial\mean{P}}{\partial x_i}\mathrm{d}t + 2\nu\frac{\partial^2\mean{U_i}}{\partial x_j\partial x_j}\mathrm{d}t + \left(2\nu\right)^{1/2}\frac{\partial\mean{U_i}}{\partial x_j}\mathrm{d}W_j\\
&\quad- \frac{1}{\rho}\frac{\partial p}{\partial x_i}\mathrm{d}t + 2\nu\frac{\partial^2u_i}{\partial x_j\partial x_j}\mathrm{d}t + \left(2\nu\right)^{1/2}\frac{\partial u_i}{\partial x_j}\mathrm{d}W_j,
\end{split}
\label{eq:Lagrangian-velocity-decomposed}
\end{equation}
the last three terms are unclosed. Closure hypotheses for these terms can be made either by suggesting approximations for the PDF fluxes (represented by the conditional expectations) in \Eqre{eq:EulerianPDFforviscous} or by proposing stochastic processes that simulate the physical phenomena represented by the second row of \Eqre{eq:Lagrangian-velocity-decomposed}. For an overview of the wide variety of modeling approaches see the review compiled by \citet{Dopazo_94}. To model the increments in particle velocity the generalized Langevin model (GLM) of \citet{Haworth_86} is adopted here
\begin{equation}
\begin{split}
\mathrm{d}\mathcal{U}_i(t)&=-\frac{1}{\rho}\frac{\partial\mean{P}}{\partial x_i}\mathrm{d}t + 2\nu\frac{\partial^2\mean{U_i}}{\partial x_j\partial x_j}\mathrm{d}t + \left(2\nu\right)^{1/2}\frac{\partial\mean{U_i}}{\partial x_j}\mathrm{d}W_j\\
&\quad+G_{ij}\left(\mathcal{U}_j-\mean{U_j}\right)\mathrm{d}t + \left(C_0\varepsilon\right)^{1/2}\mathrm{d}W'_i,
\end{split}
\label{eq:Lagrangian-model}
\end{equation}
where \(G_{ij}\) is a second-order tensor function of velocity statistics, \(C_0\) is a positive constant, \(\varepsilon\) denotes the rate of dissipation of turbulent kinetic energy and \(\mathrm{d}W'_i\) is another Wiener process. By comparing \Eqres{eq:Lagrangian-velocity-decomposed} and \Eqr{eq:Lagrangian-model}, it is apparent that the terms in \(G_{ij}\) and \(C_0\) jointly model the last three terms in \Eqre{eq:Lagrangian-velocity-decomposed} representing pressure redistribution and anisotropic dissipation of turbulent kinetic energy \citep{Pope_00}. A particular specification of \(G_{ij}\) corresponds to a particular Lagrangian stochastic model for the instantaneous particle velocity increment. Following \citet{Dreeben_98}, we specify \(G_{ij}\) and \(C_0\) employing Durbin's elliptic relaxation method using the constraint
\begin{equation}
\left(1+\tfrac{3}{2}C_0\right)\varepsilon+G_{ij}\mean{u_iu_j}=0,\label{eq:constraint-on-c0}
\end{equation}
which ensures that the kinetic energy evolves correctly in homogeneous turbulence \citep{Pope_00}. Introducing the tensor \(\wp_{ij}\) to characterize the non-local effects \(G_{ij}\) and \(C_0\) are defined as
\begin{equation}
G_{ij} = \frac{\wp_{ij}-\frac{\varepsilon}{2}\delta_{ij}}{k} \quad\mathrm{and}\quad C_0 = \frac{-2\wp_{ij}\mean{u_iu_j}}{3k\varepsilon},\label{eq:GandC}
\end{equation}
where \(k=\frac{1}{2}\mean{u_iu_i}\) represents the turbulent kinetic energy. The non-local term \(\wp_{ij}\) is specified with the following elliptic relaxation equation
\begin{equation}
\wp_{ij} - L^2\nabla^2\wp_{ij} = \frac{1-C_1}{2}k\mean{\omega}\delta_{ij} + kH_{ijkl}\frac{\partial\mean{U_k}}{\partial x_l},\label{eq:elliptic-relaxation-Lagrangian}
\end{equation}
where
\begin{equation}
\begin{split}
H_{ijkl} = (C_2A_v + \tfrac{1}{3}\gamma_5)\delta_{ik}\delta_{jl} - \tfrac{1}{3}\gamma_5\delta_{il}\delta_{jk}\\
+\gamma_5b_{ik}\delta_{jl} - \gamma_5b_{il}\delta_{jk},\label{eq:H}
\end{split}
\end{equation}
\begin{equation}
A_v = \min\left[1.0,C_v\frac{\det\mean{u_iu_j}}{\left(\frac{2}{3}k\right)^3}\right],
\end{equation}
and
\begin{equation}
b_{ij} = \frac{\mean{u_iu_j}}{u_ku_k} - \tfrac{1}{3}\delta_{ij}
\end{equation}
is the Reynolds stress anisotropy, \(\mean{\omega}\) denotes the mean characteristic turbulent frequency and \(C_1, C_2, \gamma_5, C_v\) are model constants. The characterstic lengthscale \(L\) is defined by the minimum of the turbulent and Kolmogorov lengthscales
\begin{equation}
L=C_L\max\left[C_\xi\frac{k^{3/2}}{\varepsilon},C_\eta\left(\frac{\nu^3}{\varepsilon}\right)^{1/4}\right],\label{eq:L}
\end{equation}
with
\begin{equation}
C_\xi=1.0+1.3n_in_i,\label{eq:Cxi}
\end{equation}
where \(n_i\) is the unit wall-normal of the closest wall-element pointing outward of the flow domain, while \(C_L\) and \(C_\eta\) are model constants. The definition of \(C_\xi\) in \Eqre{eq:Cxi} signifies a slight departure from the original model by attributing anisotropic and wall-dependent behavior to its value. In the case of a channel flow where the wall is aligned with \(x\), the wall-normal \(\bv{n}=(0,-1,0)\). This gives \(C_\xi=2.3\) in the computation of \(\wp_{22}\) in \Eqre{eq:elliptic-relaxation-Lagrangian} and \(C_\xi=1.0\) for all other components. The modification improves the centerline behavior of the wall-normal Reynolds stress component \(\mean{v^{\scriptscriptstyle 2}}\) and in turn the cross-stream mixing of the passive scalar. Another departure from the original model is the application of the elliptic term \(L^2\nabla^2\wp_{ij}\) (as originally proposed by \citet{Durbin_93}) as opposed to \(L\nabla^2(L\wp_{ij})\). This simplification was adopted because no visible improvement has been found by employing the second, numerically more expensive term.

The right hand side of \Eqre{eq:elliptic-relaxation-Lagrangian} may be any local model for pressure redistribution: here we follow \citet{Dreeben_98} and use the stochastic Lagrangian equivalent of a modified isotropization of production (IP) model proposed by \citet{Pope_94}. Close to the wall, the elliptic term on the left hand side of \Eqre{eq:elliptic-relaxation-Lagrangian} brings out the non-local, highly anisotropic behavior of the Reynolds stress tensor, whereas far from the wall the significance of the elliptic term vanishes and the local model on the right hand side is recovered. 

The above model needs to be augmented by an equation for a quantity that provides length-, or time-scale information for the turbulence. With traditional moment closures the most common approach is to solve a model equation for the turbulent kinetic energy dissipation rate \(\varepsilon\) itself as proposed by \citet{Hanjalic_72}. An alternative method is to solve an equation for the mean characteristic turbulent frequency \citep{Wilcox_93} \(\mean{\omega}\) and to define
\begin{equation}
\varepsilon = k\mean{\omega}.\label{eq:dissipation-from-frequency}
\end{equation}
In PDF methods, however, a fully Lagrangian description has been preferred. A Lagrangian stochastic model has been developed for the instantaneous particle frequency \(\omega\) by \citet{VanSlooten_98} of which different forms exist, but the simplest formulation can be cast into
\begin{equation}
\begin{split}
\mathrm{d}\omega = -C_3\mean{\omega}\left(\omega-\mean{\omega}\right)\mathrm{d}t - S_\omega\mean{\omega}\omega\mathrm{d}t\\
+\left(2C_3C_4\mean{\omega}^2\omega\right)^{1/2}\mathrm{d}W,\label{eq:frequency-model}
\end{split}
\end{equation}
where \(S_\omega\) is a source/sink term for the mean turbulent frequency
\begin{equation}
S_\omega=C_{\omega2}-C_{\omega1}\frac{\mathcal{P}}{\varepsilon},\label{eq:frequency-source}
\end{equation}
where \(\mathcal{P}=-\mean{u_iu_j}\partial\mean{U_i}/\partial x_j\) is the production of turbulent kinetic energy, \(\mathrm{d}W\) is a scalar valued Wiener-process, while \(C_3,C_4,C_{\omega1}\) and \(C_{\omega2}\) are model constants. Since the no-slip condition would incorrectly force \(\varepsilon\) to zero at the wall, \Eqre{eq:dissipation-from-frequency} needs to be modified, thus the dissipation is defined as \citep{Dreeben_98}
\begin{equation}
\varepsilon = \mean{\omega}\left(k + \nu C_T^2\mean{\omega}\right),\label{eq:dissipation-from-frequency_wall}
\end{equation}
where \(C_T\) is also a model constant. A simplification of the original model for the turbulent frequency employed by \citet{Dreeben_98} is the elimination of the ad-hoc source term involving an additional constant, since no obvious improvement has been found by including it. 

Similarly to the model equations for the Lagrangian velocity and frequency increments, the evolution equations of the passive scalar are also given in Lagrangian form, which define the two micromixing models that are investigated in this study:
\begin{eqnarray}
\mathrm{d}\psi&=&-\frac{1}{t_\mathrm{m}}\left(\psi-\mean{\phi}\right)\mathrm{d}t\qquad\quad\mathrm{(IEM)},\label{eq:IEM}\\
\mathrm{d}\psi&=&-\frac{1}{t_\mathrm{m}}\left(\psi-\mean{\phi|\bv{V}}\right)\mathrm{d}t\qquad\mathrm{(IECM)},\label{eq:IECM}
\end{eqnarray}
where \(\psi\) represents the sample space variable of the species concentration \(\phi\), \(t_\mathrm{m}\) is the micromixing timescale, while \(\mean{\phi|\bv{V}}=\mean{\phi|\bv{U}(\bv{x},t)=\bv{V}}\) denotes the expected value of the mean concentration conditional on the velocity. Both of these models represent the physical process of dissipation and reflect the concept of relaxation towards a scalar mean with the characteristic timescale \(t_\mathrm{m}\). The difference is that in the IEM model, all particles that have similar position interact with each other, while in the IECM model only those particles interact that also have similar velocities, e.g.\ fluid elements that belong to the same eddy.

It can be shown that in the case of homogeneous turbulent mixing with no mean scalar gradient the two models are equivalent \citep{Fox_96} and the micromixing timescale \(t_\mathrm{m}\) is proportional to the Kolmogorov timescale \(\tau=k/\varepsilon\). In the inhomogeneous case of a concentrated source, however, there are various stages of the spreading of the plume requiring different characterizations of \(t_\mathrm{m}\).  In this case, the formal simplicity of the IEM and IECM models is a drawback, since a single scalar parameter \(t_\mathrm{m}\) has to account for all the correct physics. The timescale should be inhomogeneous and should depend not only on the local turbulence characteristics but also on the source location, type, size, distribution and strength. Because of this complexity, a general flow-independent specification of \(t_\mathrm{m}\) has been elusive. In the following, we define a micromixing timescale for a passive scalar released from a concentrated source into an inhomogeneous flow surrounded by no-slip walls.

In general, \(t_\mathrm{m}\) is assumed to be proportional to the timescale of the instantaneous plume \citep{Sawford_04}. Once the initial conditions are forgotten, theoretical results \citep{Franzese_07} show that the timescale of the instantaneous plume is linear in \(t\) in the inertial subrange and is proportional to the turbulence timescale in the far field, when the instantaneous plume grows at the same rate as the mean plume. Based on these considerations the micromixing timescale is computed according to
\begin{equation}
t_\mathrm{m} = \min\left[C_s\left(\frac{r_0^2}{\varepsilon}\right)^{1/3} + C_t\frac{x}{\mean{U}_c};\enskip\max\left(\frac{k}{\varepsilon}, C_T\sqrt{\frac{\nu}{\varepsilon}}\right)\right],\label{eq:micromixing-timescale}
\end{equation}
where \(r_0\) denotes the radius of the source, \(\mean{U}_c\) is the mean velocity at the centerline of the channel, while \(C_s\) and \(C_t\) are micromixing model constants. This definition reflects the three stages of the spreading of the plume. In the first stage, the time scale of the plume is proportional to that of the source \citep{Batchelor_52}: accordingly, the first term in the \(min\) operator represents the effect of the source. In the second stage \(t_\mathrm{m}\) increases linearly as the scalar is dispersed downstream and the distance \(x\) from the source grows \citep{Franzese_07}. In the final stage, the timescale is capped with the characteristic timescale of the turbulence, which provides an upper limit in the third term of \Eqre{eq:micromixing-timescale}. Following \citet{Durbin_91} this is defined as the maximum of the turbulent and Kolmogorov timescales: far from the boundaries it becomes \(k/\varepsilon\), whereas near a surface, where \(k \rightarrow 0\), the Kolmogorov timescale provides a lower bound as \(C_T(\nu/\varepsilon)^{1/2}\).

The equations to model the joint PDF of velocity and turbulent frequency closely follow the work of \citet{Dreeben_98}. Slight modifications consist of
\begin{itemize}
\item the anisotropic definition of lengthscale \(L\) in \Eqres{eq:L} and \Eqr{eq:Cxi},
\item the application of the elliptic term \(L^2\nabla^2\wp_{ij}\) instead of \(L\nabla^2(L\wp_{ij})\) in \Eqre{eq:elliptic-relaxation-Lagrangian}, and
\item the elimination of an ad-hoc source term in \Eqre{eq:frequency-source}.
\end{itemize}
These modifications do not make the methodology less general or flow-dependent, i.e.\ the velocity model remains complete. On the other hand, the specification of \(t_\mathrm{m}\) in \Eqre{eq:micromixing-timescale} is somewhat flow-dependent, since we assume that the scalar plume will be dispersed in the \(x\) (downstream) direction.

In summary, the flow is represented by a large number of Lagrangian particles representing a finite sample of all fluid particles in the domain. Each particle has position \(\mathcal{X}_i\) and carries its velocity \(\mathcal{U}_i\), turbulent frequency \(\omega\) and concentration \(\psi\). These particle properties are advanced according to \Eqres{eq:Lagrangian-position-exact}, \Eqr{eq:Lagrangian-model}, \Eqr{eq:frequency-model} and either \Eqr{eq:IEM} or \Eqr{eq:IECM}, respectively. The discretized particle equations are advanced in time by the first order accurate forward Euler method. This method was preferred to the more involved exponential scheme that was originally suggested by \citet{Dreeben_98}, since the code is sufficiently stable with the simpler and computationally less expensive Euler method as well.

\section{Numerical method}
\label{sec:num_method}
In turbulent channel flow after an initial development region, the statistics of fluid dynamics (e.g.\ turbulent velocity and frequency) are expected to become homogeneous in the streamwise direction, while strongly inhomogeneous in the wall-normal direction. A passive scalar released into this flow from a concentrated source initially exhibits high inhomogeneity and only far downstream becomes fully mixed. We will separately discuss the numerical issues related to the streamwise statistically homogeneous one-dimensional fluid dynamics and the inhomogeneous two-dimensional scalar field.

\subsection{Modeling the fluid dynamics}
While the velocity and turbulent frequency statistics become one-dimensional, both the streamwise \(x\) and cross-stream \(y\) components of the particle position are retained so that particles can represent the streamwise inhomogeneity of the scalar. A one-dimensional grid, that is refined at the wall, is used to compute Eulerian statistics of the velocity and turbulent frequency by ensemble averaging in elements. The elliptic-relaxation equation \Eqr{eq:elliptic-relaxation-Lagrangian} is also solved on this grid with a finite element method. A constant unit mean streamwise pressure gradient is imposed, drives the flow and builds up the numerical solution. The cross-stream mean pressure gradient is obtained by satisfying the cross-stream mean momentum equation
\begin{equation}
\frac{1}{\rho}\frac{\mathrm{d}\mean{P}}{\mathrm{d}y} = -\frac{\mathrm{d}\mean{v^{\scriptscriptstyle 2}}}{\mathrm{d}y}.
\end{equation}
Wall-boundary conditions for the particles are the same as suggested by \citet{Dreeben_98} but repeated here for clarity. Over any given time-interval a particle undergoing reflected Brownian motion in the vicinity of a wall may strike the wall infinitely many times \citep{Pope_00}. Therefore wall-conditions have to be enforced on particles that either \emph{penetrate} or \emph{possibly penetrate} the wall during timestep \(\Delta t\), the probability of which can be calculated by \citep{Karatzas_91}
\begin{equation}
P=\exp\left(\frac{-\mathcal{Y}_0\mathcal{Y}}{\nu\Delta t}\right),
\end{equation}
where \(\mathcal{Y}_0\) and \(\mathcal{Y}\) denote the distance of the particle from the wall at the previous and current timestep, respectively. Thus particle wall-conditions are applied if
\begin{equation}
\mathcal{Y} < 0,
\end{equation}
or if
\begin{equation}
\mathcal{Y} \geqslant 0 \quad \mathrm{and} \quad \exp\left(\frac{-\mathcal{Y}_0\mathcal{Y}}{\nu\Delta t}\right) > \eta,
\end{equation}
where \(\eta\) is a random variable with a standard uniform distribution. The imposed wall-conditions are
\begin{equation}
\mathcal{U}_i = 0 \qquad \textrm{(no slip)}
\end{equation}
and the particle frequency is sampled from a gamma distribution with mean and variance
\begin{equation}
\mean{\omega} = \frac{1}{C_T}\frac{\mathrm{d}\sqrt{2k}}{\mathrm{d}y} \quad \mathrm{and} \quad C_4\mean{\omega}^2.\label{eq:frequency-wallstats}
\end{equation}
Only half of the channel is retained for the purpose of modeling the fluid dynamics. The particle conditions for the centerline are symmetry conditions, i.e.\ if the particle tries to leave the domain through the centerline, it is reflected back with opposite normal velocity. Consistently with the above particle conditions, boundary conditions are imposed on the Eulerian statistics as well. At the wall, the mean velocity and the Reynolds stress tensor are forced to zero. The mean frequency \(\mean{\omega}\) is set according to \Eqre{eq:frequency-wallstats}. At the centerline, the shear Reynolds stress \(\mean{uv}\) is set to zero. At the wall in the elliptic-relaxation equation \Eqr{eq:elliptic-relaxation-Lagrangian}, \(\wp_{ij}\) is set according to \(\wp_{ij}=-4.5\varepsilon n_in_j\). In the current case the wall is aligned with \(y=0\), thus only the wall-normal component is non-zero: \(\wp_{22}=-4.5\varepsilon\). At the centerline, symmetry conditions are applied for \(\wp_{ij}\), i.e.\ homogeneous Dirichlet-conditions are applied for the off-diagonal components and homogeneous Neumann-conditions for the diagonal components.

An equal number of particles are uniformly generated into each element and initial particle velocities are sampled from a Gaussian distribution with zero mean and variance 2/3, i.e.\ the initial Reynolds stress tensor is isotropic with unit turbulent kinetic energy. Initial particle frequencies are sampled from a gamma distribution with unit mean and variance 1/4.

\subsection{Modeling the passive scalar}
\label{sec:scalar-modeling}
A passive, inert scalar is released from a concentrated source. The scalar statistics are inhomogeneous and in general not symmetric about the channel centerline, thus a second, two-dimensional grid is employed to calculate scalar statistics. The use of separate grids for the fluid dynamics and scalar fields enables the grid refinement to be concentrated on different parts of the domain, i.e.\ the scalar-grid can be refined around the source, while the fluid dynamics-grid is refined at the wall. The two-dimensional grid is unstructured and consists of triangles. The role of this mesh is twofold: it is used for computing Eulerian scalar statistics and for tracking particles on the domain. Since the scalar is passive, only one-way coupling between the two grids is necessary. This is accomplished by using the local velocity statistics computed in the 1d-elements for those 2d-elements which lie closest to them in the wall-normal coordinate direction. The Eulerian statistics in both grids are computed by a two-step procedure: first ensemble averaging is used to compute statistics in elements, then these element-based statistics are transferred to nodes by averaging the elements surrounding the nodes. When Eulerian statistics are needed in particle equations, the average of the nodal values are used for all the particles that reside in the given element. Spatial derivatives are computed with linear finite element shapefunctions for triangles \citep{Lohner_01}. The long two-dimensional rectangular domain is also subdivided into several equally sized bins. The velocity and turbulent frequency statistics are computed using the one-dimensional grid in which only particles in the first bin participate. The position of these particles are then copied to all downstream bins and mirrored to the upper half of the channel. If the particle hits the centerline, its concentration is exchanged with its mirrored pair on the upper half, allowing a possible non-symmetric behavior of the scalar.

In order to numerically compute the expected value of the mean concentration conditional on the velocity field \(\mean{\phi|\bv{V}}\) in \Eqre{eq:IECM}, one needs to discretize the velocity sample space \(\bv{V}\) and compute different means for each sample space bin. A straightforward way to implement this is to equidistantly divide the three-dimensional velocity space into cubical bins and compute separate scalar means for each cube using those particles whose velocity fall into the given cube. A better way of choosing the conditioning intervals for each dimension is to define their endpoints so that the probabilities of the particle velocities falling into the bins are equal. For a Gaussian velocity PDF these endpoints can be obtained from statistical tables as suggested by \citet{Fox_96} for homogeneous flows.  If the approximate shape of the joint velocity PDF is not known, as in our case, a different sort of algorithm is required to homogenize the statistical error over the sample space.

Note that the sample-spatial distribution of the conditioning intervals has to be neither equidistant nor the same in all three dimensions and can also vary from element to element. Therefore, we determine the binning dynamically so that no bin remains without particles and the number of particles in each bin is approximately the same. A judicious sorting and grouping of the particles according to their velocity components can be used to achieve this. We compute conditioned means in each triangular element as follows. For a binning of \(2\times2\times2\) (i.e.\ the desired total number of bins is 8), first the particles residing in the triangle are sorted according to their \(\mathcal{U}\) velocity component. Then both the first and the second halves of the group are separately sorted according to their \(\mathcal{V}\) velocity component. After further dividing both halves into halves again, the quarters are sorted according to the \(\mathcal{W}\) component. Finally, halving the four quarters into eight subgroups, we compute scalar means for each of these eight subgroups. This procedure is general, it employs no assumptions about the shape of the velocity PDF and homogenizes the statistical error over the sample space. It also ensures that there will be no empty bins as long as the number of conditioning bins and the number of particles/elements are reasonable. In principle, the number of subdivisions (i.e.\ the sample-spatial refinement) can be arbitrary. We employ the binning structure of \(4\times4\times4\) throughout the current calculations.

In order to reduce the statistical error during the computation of \(\mean{\phi|\bv{V}}\), \citet{Fox_96} proposed an alternative method in which the three dimensional velocity space is projected onto a one dimensional sub-space where the sample-spatial discretization is carried out. This way, a relatively large number of particles can be used to obtain a local \(\mean{\phi|\bv{V}}\), which results in a smaller statistical error. The projection however is exact only in the case of a passive inert scalar in homogeneous turbulent shear flows with a uniform mean scalar gradient. In the current, more general inhomogeneous context, this projection could still be used as a modeling assumption \citep{Fox_96}. To explore the error introduced by the projection in modeling inhomogeneous flows, we implemented and tested this method for the channel flow. For both releases (to be discussed in the next sections) we found that the projection has the largest influence on the scalar PDFs by homogenizing their high peaks, i.e.\ shaping them closer to a Gaussian. In the case of the centerline release, an artifact of double-peaks in the r.m.s.\ profiles of the cross-stream distribution of the passive scalar can also be attributed solely to this projection method. In the following sections, only the former, more general, three-dimensional method was used in conjunction with the IECM model with \(4^3\) bins.

\section{Results}
\label{sec:results}
The model has been run for the case of fully developed channel flow at \(Re_\tau=1080\) based on the friction velocity \(u_\tau\) and the channel half-width \(h\) with a passive scalar released from a concentrated source at the centerline \((y_s/h=1.0)\) and in the viscous wall region \((y_s/h=0.067)\). The results are divided into a discussion of the fluid dynamics statistics (\ref{sec:res_fd}), a comparison of the two micromixing models (\ref{sec:res_comp}) and a presentation of unconditional (\ref{sec:res_stat}) and conditional (\ref{sec:res_cond_stat}) scalar statistics.

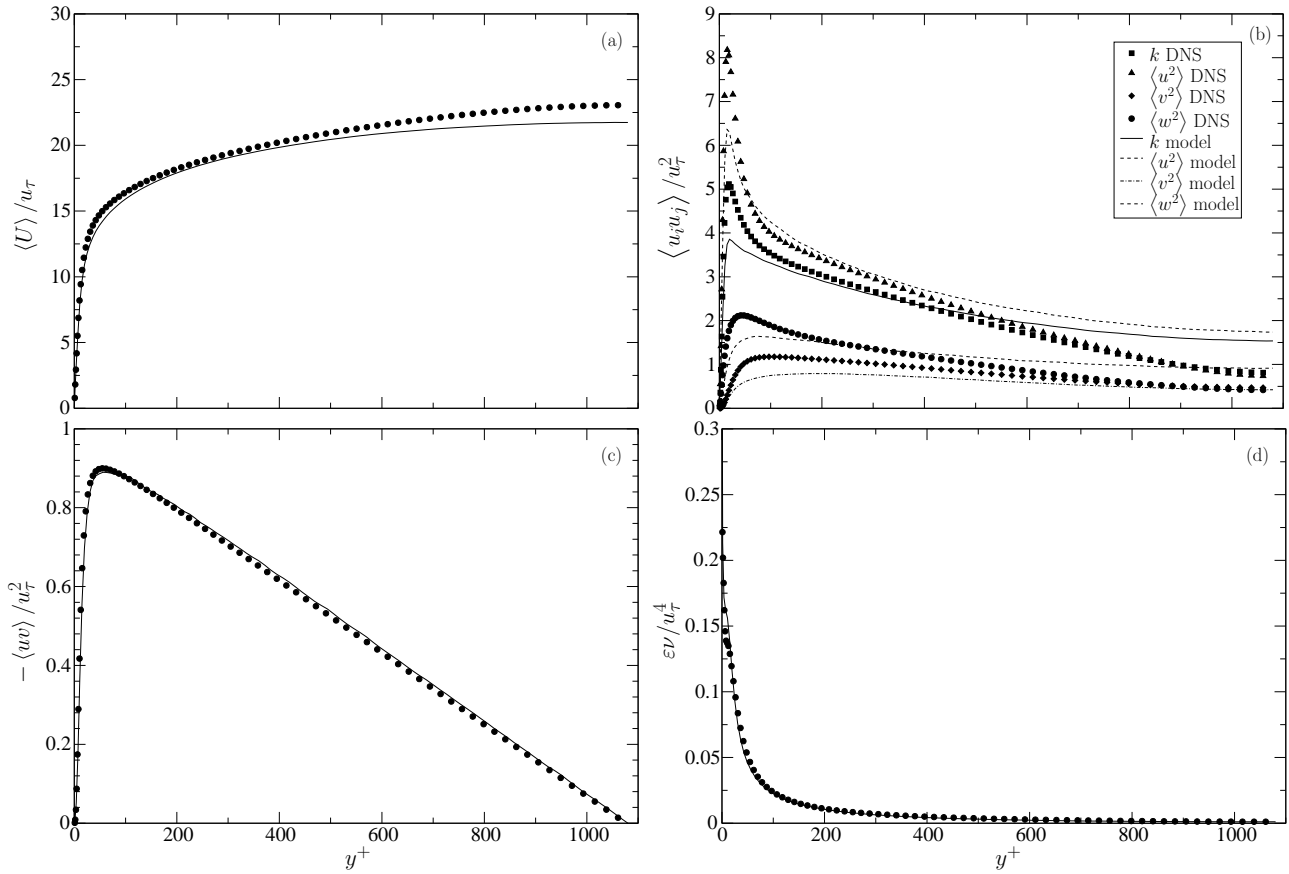
\begin{figure*}
\centering
\resizebox{17cm}{!}{\input{velocity.pstex_t}}
\caption{\label{fig:velocity}Cross-stream profiles of (a) the mean streamwise velocity, (b) the diagonal components of the Reynolds stress tensor, (c) the shear Reynolds stress and (d) the rate of dissipation of turbulent kinetic energy. Lines -- PDF calculation, symbols -- DNS data of \citet{Abe_04}. All quantities are normalized by the friction velocity and the channel half-width. The DNS data is scaled from \(Re_\tau=1020\) to \(1080\).}
\end{figure*}
\subsection{Fluid dynamics}
\label{sec:res_fd}
\begin{table}
\caption{\label{tab:fd-constants}Constants for modeling the joint PDF of velocity and frequency.}
\begin{ruledtabular}
\begin{tabular}{ccccccccccc}
\(C_1\)&\(C_2\)&\(C_3\)&\(C_4\)&\(C_T\)&\(C_L\)&\(C_\eta\)&\(C_v\)&\(\gamma_5\)&\(C_{\omega1}\)&\(C_{\omega2}\)\\
\hline
1.85&0.63&5.0&0.25&6.0&0.134&72.0&1.4&0.1&0.5&0.73
\end{tabular}
\end{ruledtabular}
\end{table}
 The equations to model the velocity and turbulent frequency have been solved on a 100-cell one dimensional grid with 500 particles per cell. The applied model constants are displayed in Table \ref{tab:fd-constants}. The computed cross-stream profiles of mean streamwise velocity, the non-zero components of the Reynolds stress tensor and the rate of dissipation of turbulent kinetic energy are compared with the DNS data of \citet{Abe_04} at \(\textit{Re}_\tau=1020\) in \Fige{fig:velocity}. Previous PDF modeling studies employing the elliptic relaxation technique \citep{Dreeben_97,Dreeben_98,Waclawczyk_04} have been conducted up to \(\textit{Re}_\tau=590\). The high-level inhomogeneity and anisotropy in the viscous wall region are well represented by the technique at this higher Reynolds number as well. The purpose of including the parameter \(C_\xi\) in \Eqre{eq:L} of the wall-normal component of \(\wp_{ij}\) is to correct the overprediction of the wall-normal Reynolds stress component \(\mean{v^{\scriptscriptstyle 2}}\) at the centerline. This facilitates the correct behavior of the mean of the dispersed passive scalar in the center region of the channel (presented in Section \ref{sec:res_comp}).

\subsection{Comparison of the IEM and IECM micromixing models}
\label{sec:res_comp}
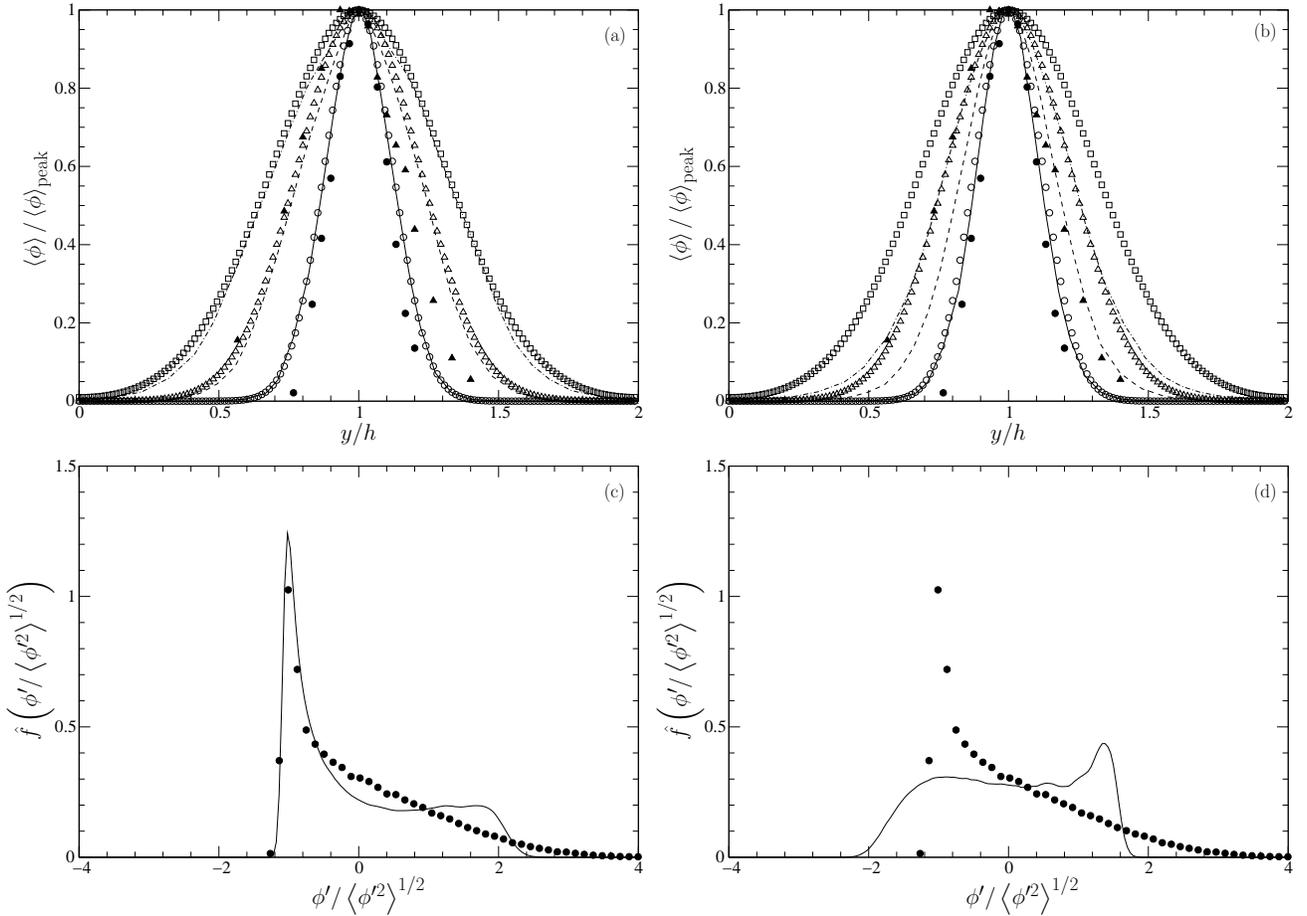
\begin{figure*}
\centering
\resizebox{17cm}{!}{\input{comparison.pstex_t}}
\caption{\label{fig:comparison}Cross-stream mean concentration profiles normalized by their respective peak values at different downstream locations as computed by the (a) IECM and (b) IEM models for the centerline release. Lines -- PDF calculation at solid line, \(x/h=4.0\), dashed line, \(x/h=7.4\) and dot-dashed line, \(x/h=10.8\), hollow symbols -- analytical Gaussians using \Eqre{eq:Gaussians} at \(\circ\),~\(x/h=4.0\); \(\vartriangle\),~\(x/h=7.4\) and \(\Box\),~\(x/h=10.8\), filled symbols -- experimental data of \citet{Lavertu_05} at \(\bullet\),~\(x/h=4.0\) and \(\blacktriangle\),~\(x/h=7.4\).  Also shown, PDFs of scalar concentration fluctuations at (\(x/h=7.4\), \(y/h=1.0\)) for the (c) IECM and (d) IEM models. Lines -- computation, symbols -- experimental data.}
\end{figure*}
An often raised criticism of the IEM model is that there is no physical basis for assuming the molecular mixing to be independent of the velocity field. This assumption gives rise to a spurious (and unphysical) source of scalar flux \citep{Pope_98}. This behavior of the IEM model has also been demonstrated for line sources in homogeneous grid turbulence \citep{Sawford_04}. The situation can be remedied by introducing the velocity-conditioned scalar mean \(\mean{\phi|\bv{V}}\), which leads to the IECM model. Often invoked as a desirable property of micromixing models is that the scalar PDF should tend to a Gaussian  for homogeneous turbulent mixing \citep{Pope_00,Fox_03} (i.e.\ statistically homogeneous scalar field in homogeneous isotropic turbulence).  While mathematically a Gaussian does not satisfy the boundedness property of the advection-diffusion scalar transport equation, it is generally assumed that the limiting form of the PDF can be reasonably approximated by a clipped Gaussian.  Also, \citet{Chatwin_02,Chatwin_04} argued that in most practical cases, where the flow is inhomogeneous, the scalar PDF is better approximated by non-Gaussian functions, which should ultimately converge to a Dirac delta function about the mean, \(\delta(\psi-\mean{\phi})\), where \(\mean{\phi}\) approaches a positive value in bounded domains and zero in unbounded domains. 

In fully developed turbulent channel flow the center region of the channel may be considered approximately homogeneous \citep{Brethouwer_01,Vrieling_03}. Thus for a centerline source, up to a certain downstream distance where the plume still lies completely in the center region, the mean scalar field can be described by Taylor's theory of absolute dispersion \citep{Taylor_21}. Likewise, numerical simulations are expected to reproduce experimental measurements of grid turbulence. According to the theory, the mean-square particle displacement \(\mean{\mathcal{Y}^2}\) is related to the autocorrelation function of the Lagrangian velocity \(R_L=\mean{v(t)v(t')}/\mean{v^{\scriptscriptstyle{2}}}\) as
\begin{equation}
\mean{\mathcal{Y}^2} = 2\mean{v^2}\int_0^t\int_0^{t'}R_L(\xi)\mathrm{d}\xi\mathrm{d}t',\label{eq:particle-displacement-Taylor}
\end{equation}
where it is assumed that in stationary turbulence \(R_L\) depends only on the time difference \(\xi=t-t'\). Lagrangian statistics such as \(R_L(\xi)\) are difficult to determine experimentally. An analytical expression that is consistent with the theoretically predicted behavior of the Lagrangian spectrum in the inertial subrange is \citep{Arya_99}
\begin{equation}
R_L(\xi)=\exp\left(-\frac{|\xi|}{T_L}\right),\label{eq:expR}
\end{equation}
where \(T_L\) denotes the Lagrangian integral timescale. Substituting \Eqre{eq:expR} into \Eqre{eq:particle-displacement-Taylor} the following analytical expression can be obtained for the root-mean-square particle displacement
\begin{equation}
\sigma^2_y=\mean{\mathcal{Y}^2}=2\mean{v^2}T^2_L\left[\frac{t}{T_L}-1+\exp\left(-\frac{t}{T_L}\right)\right].
\end{equation}
This expression can be used to approximate the spread of the plume that is released at the centerline of the channel. As Lagrangian timescale we take
\begin{equation}
T_L=\frac{2\mean{v^{\scriptscriptstyle 2}}}{C_0\varepsilon},
\end{equation}
where \(C_0\) is usually taken as the Lagrangian velocity structure function inertial subrange constant \citep{Monin_Yaglom_75,Sawford_06}, which ensures consistency of the Langevin equation \Eqr{eq:Lagrangian-model} with the Kolmogorov hypothesis in stationary isotropic turbulence \citep{Pope_00}. In the current case the value of \(C_0\) is defined by \Eqre{eq:GandC} and is no longer a constant, but depends on the velocity statistics. For the purpose of the current analytical approximation, however, a constant value (0.8) has been estimated as the spatial average of \(C_0\) computed by \Eqre{eq:GandC}. For the cross-stream Reynolds stress \(\mean{v^{\scriptscriptstyle 2}}\) and the dissipation rate \(\varepsilon\) their respective centerline values are employed. In analogy with time \(t\) in homogeneous turbulence, we define \(t=x/\mean{U}_c\), where \(x\) is the downstream distance from the source and \(\mean{U}_c\) is the mean velocity at the centerline. Thus the cross-stream mean scalar profiles predicted by \Eqre{eq:particle-displacement-Taylor} are obtained from the Gaussian distribution
\begin{equation}
\mean{\phi(y)}=\frac{Q}{\mean{U}_c\left(2\pi\sigma_y^2\right)^{1/2}}\exp\left[-\frac{\left(y-y_s\right)^2}{2\sigma_y^2}\right],\label{eq:Gaussians}
\end{equation}
where \(Q\) is the source strength and \(y_s\) is the cross-stream location of the source.

\begin{table}
\caption{\label{tab:micromixing-constants}Model constants of the micromixing timescale \(t_\mathrm{m}\) defined by \Eqre{eq:micromixing-timescale} for both the IEM and IECM models.}
\begin{ruledtabular}
\begin{tabular}{rllcc}
\multicolumn{3}{c}{source location}&\(C_s\)&\(C_t\)\\
\hline
centerline&\(y_s/h=1.0\)&\(y^+=1080\)&0.02&0.7\\
wall&\(y_s/h=0.067\)&\(y^+=72\)&1.5&0.001\\
\end{tabular}
\end{ruledtabular}
\end{table}
After the velocity field converged to a statistically stationary state, a passive scalar is continuously released from a concentrated source. Two release cases have been investigated, where the scalar has been released at the centerline (\(y_s/h=1.0\)) and in the close vicinity of the wall (\(y_s/h=0.067\)). The viscous wall region experiences the most vigorous turbulent activity. The turbulent kinetic energy, its production and its dissipation and the level of anisotropy all experience their peak values in this region, see also \Fige{fig:velocity} (b). This suggests a significantly different level of turbulent mixing between the two release cases. Accordingly, the constants that determine the behavior of the micromixing timescales have been selected differently. Both the IEM and IECM models have been investigated with the micromixing timescale defined by \Eqre{eq:micromixing-timescale} using the model constants displayed in Table \ref{tab:micromixing-constants}.

The different behavior of the two models is demonstrated in \Fige{fig:comparison}, which shows mean concentration profiles for the centerline release computed by both the IEM and IECM models together with the analytical Gaussian solution (\ref{eq:Gaussians}) and the experimental data of \citet{Lavertu_05} for turbulent channel flow. Indeed, the downstream evolution of the cross-stream mean concentration profiles computed by the IECM model follows the Gaussians and is expected to deviate far downstream in the vicinity of the walls, where the effect of the walls is no longer negligible. It is also apparent in \Fige{fig:comparison} (b) that the IEM model changes the mean concentration, as expected. As discussed by \citet{Lavertu_05}, the measurements of the mean concentration experience the largest uncertainty due to inaccuracies in estimating the free-stream mean. Also, to improve the signal-to-noise ratio far downstream, a thicker wire had to be employed for measurements performed on the second half of the length considered, i.e.\ \(x/h>11.0\). These difficulties are probably the main source of the discrepancy between the experimental data and the agreeing analytical and numerical results for the case of the centerline release. Because of these inconsistencies only results for the first half of the measured channel length \((x/h<11.0)\) are considered in the current study.

The marginal PDF of scalar concentration can be obtained from the joint PDF \(f(\bv{V},\psi)\) by integrating over the velocity space
\begin{equation}
\Hat{f}(\psi)=\int f(\bv{V},\psi)\mathrm{d}\bv{V}.\label{eq:marginal-scalar-pdf}
\end{equation}
According to experimental data in grid turbulence \citep{Sawford_04} the skewness at the centerline is expected to be negative close to the source and to become positive only farther downstream. At \(x/h=7.4\), \(y/h=1.0\) the temperature PDF measured by \citet{Lavertu_05} suggests positive skewness in accordance with Sawford's data \citep{Sawford_04}. In \Fige{fig:comparison} (c) and (d) the normalized PDFs of scalar concentration fluctuations at this location as computed by both models are depicted. As opposed to the IEM model prediction, both the location of the peak and the overall shape of the PDF are captured correctly by the IECM model.

The different behavior of the two micromixing models is apparent in all one point statistics considered, with the IECM model producing a closer agreement to experimental data. The price to pay for the higher accuracy is an additional 30-40\% in CPU time as compared to the IEM model. In the remaining section only the IECM model results are considered.

\subsection{Scalar statistics with the IECM model}
\label{sec:res_stat}
\begin{figure*}
\centering
\resizebox{15.2cm}{!}{\input{cross_stream_stats_condensed.pstex_t}}
\caption{\label{fig:cross-stream-stats}Cross-stream distributions of the first four moments of scalar concentration at different downstream locations for (a)--(d) the centerline release (\(y_s/h = 1.0\)) and (e)--(h) the wall release (\(y_s/h = 0.067\)). Lines -- calculations, symbols -- experimental data at solid line, \(\bullet\), \(x/h=4.0\); dashed line, \(\blacktriangle\), \(x/h=7.4\) and dot-dashed line, \(\blacksquare\),~\(x/h=10.8\). The horizontal dashed lines for the skewness and kurtosis profiles indicate the Gaussian values of 0 and 3, respectively. Note the logarithmic scale of the kurtosis profiles.}
\end{figure*}
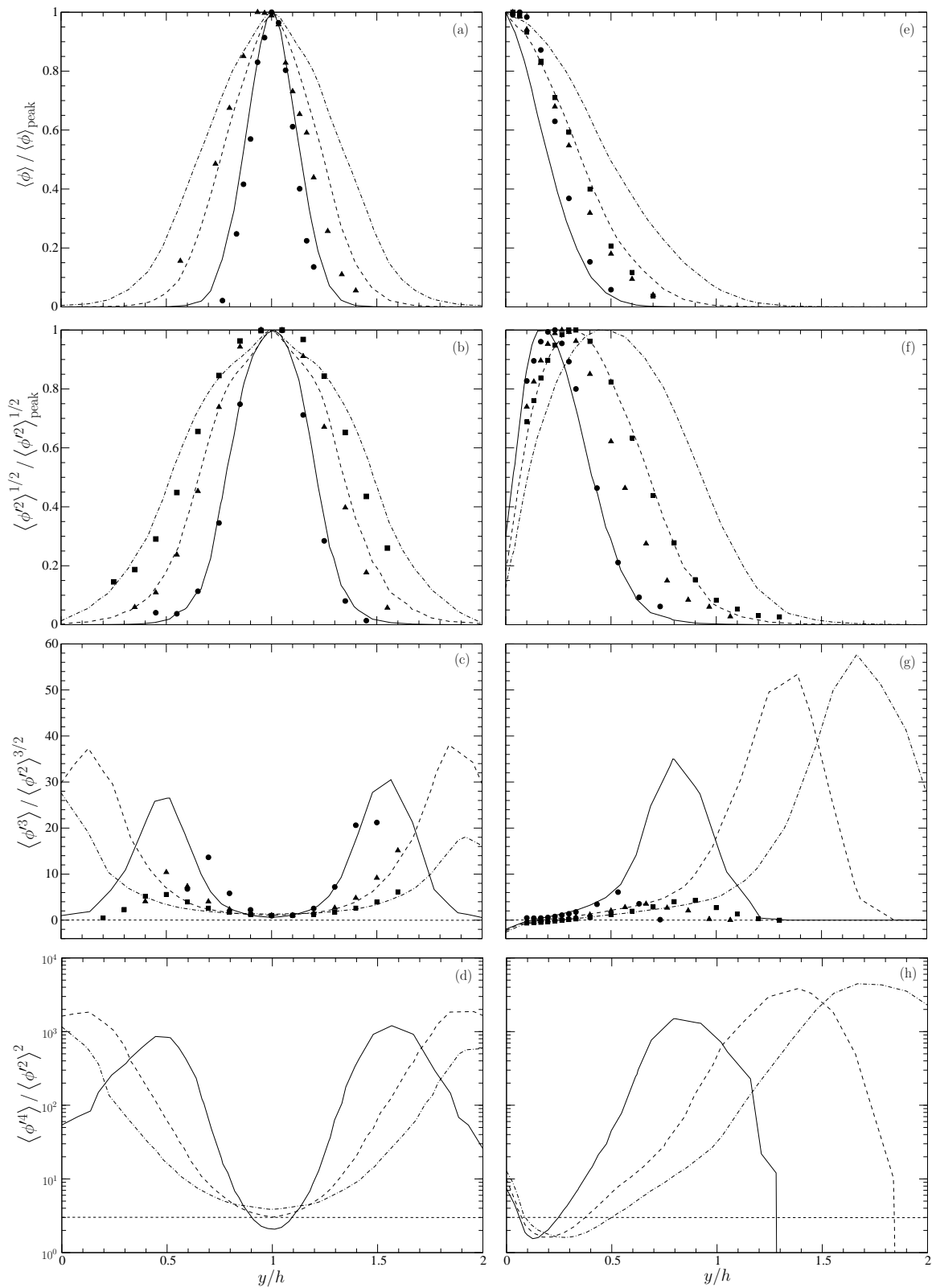
Cross-stream distributions of the first four moments of the scalar concentration at different downstream locations are shown in \Fige{fig:cross-stream-stats} for both release scenarios. The results are compared to experimental data where available.

The mean and root-mean-square (r.m.s.) profiles are normalized by their respective peak values. The width of the mean concentration profiles is most affected by the wall-normal Reynolds stress component \(\mean{v^{\scriptscriptstyle 2}}\) which is responsible for cross-stream mixing. Due to the underprediction of this component by the velocity model throughout most of the inner layer \((y^+<800)\) and the uncertainties in the experimental data mentioned in Section \ref{sec:res_comp}, the mean concentration profiles in \Fige{fig:cross-stream-stats} should be considered at most qualitative.

For the wall-release, the r.m.s.\ profiles display a clear drift of the peaks towards the centerline with increasing distance from the source \Fige{fig:cross-stream-stats} (f). This tendency has also been observed in turbulent boundary layers by \citet{Fackrell_82} and \citet{Raupach_83}. Since the scalar is statistically symmetric, in the case of the centerline release, no tranverse drift of the r.m.s.\ profiles is expected, \Fige{fig:cross-stream-stats} (b). Double peaking of the r.m.s.\ profiles has been observed in homogeneous turbulence by \citet{Warhaft_00} and \citet{Karnik_89}, noting that the profiles are initially double-peaked close to the source, then single-peaked for a short distance and then again double-peaked far downstream. \citet{Lavertu_05} found no double peaks in their measurements. Corresponding to the channel flow experiments, the PDF simulation exhibits no double-peaking in the r.m.s.\ profiles. Applying the projection method to compute \(\mean{\phi|\bv{V}}\) as described in Section \ref{sec:scalar-modeling} results in double peaking of the r.m.s.\ profiles, which is possibly due to a loss of statistical information due to its Gaussian assumption of the velocity PDF.

Skewness profiles are depicted in \Fige{fig:cross-stream-stats} (c) and (g). For both release cases, near the centers of the plumes the skewness is close to zero, indicating that the PDFs of the scalar concentration downstream of the sources are approximately symmetric. Towards the edges of the plumes however, the PDFs become very highly positively skewed, with a sudden drop to zero in the skewness outside of the plume. As observed by \citet{Lavertu_05}, the downstream evolutions of the skewness profiles indicate the eventual mixing of the plume, with the high peaks decreasing. In the current simulations the high skewness-peaks at the edge of the plumes start increasing first to even higher levels (up to about \(x/h=10.0\)) and only then start decreasing. In the case of the wall-release, the negative skewness in the viscous wall region (also apparent in the experimental data) becomes even more pronounced in the buffer layer and in the viscous sublayer, where experimental data is no longer available. The kurtosis values are close to the Gaussian value of 3 at the cross-stream location of the sources, but show significant departures towards the edges of the plume.

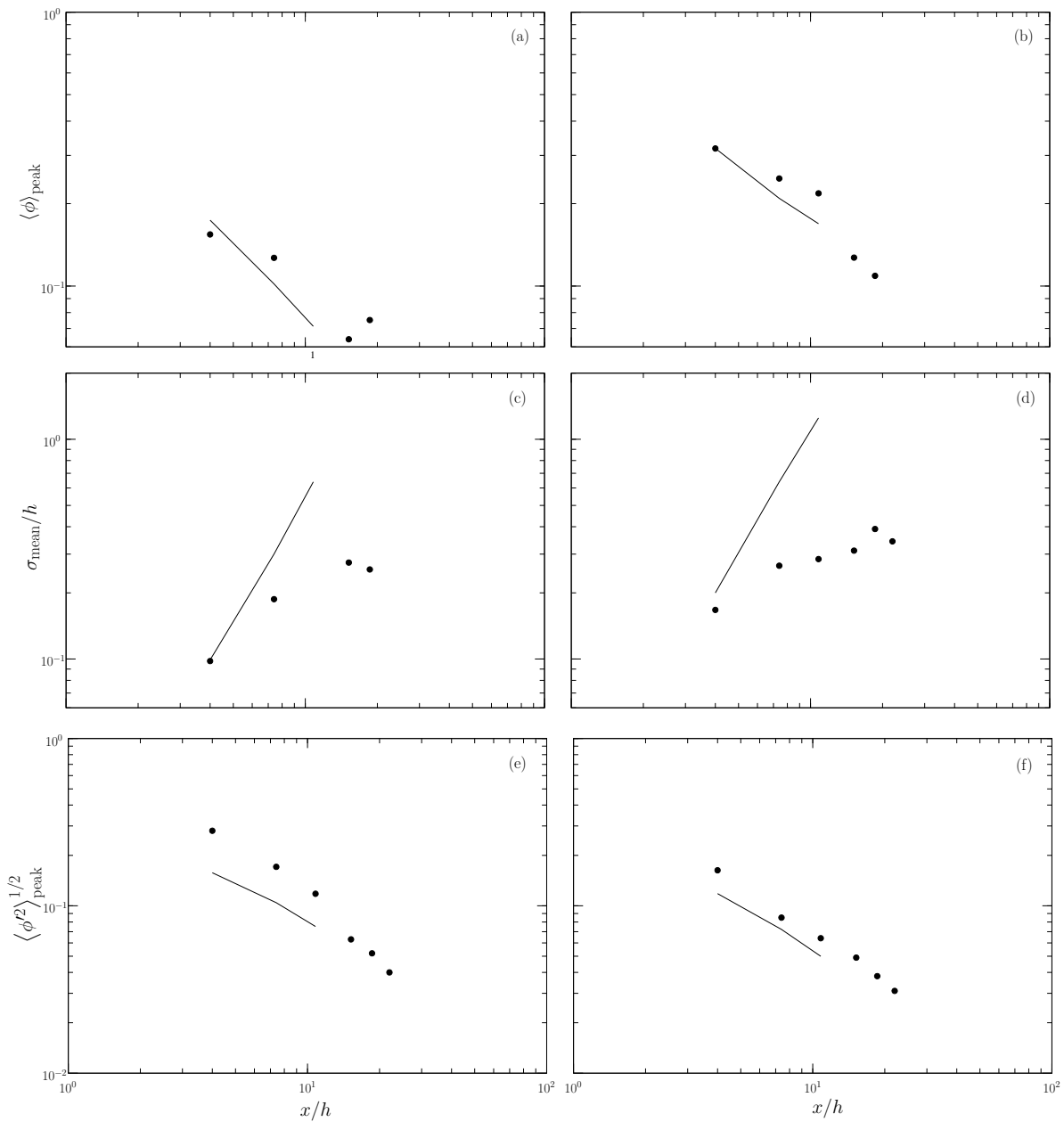
\begin{figure*}
\centering
\resizebox{15.5cm}{!}{\input{downstream_evolutions_condensed.pstex_t}}
\caption{\label{fig:downstream-evolutions}Downstream evolutions of (a), (b) the peak mean scalar concentration, (c), (d) the width of the mean concentration and (e), (f) the peak of the r.m.s.\ profiles for the centerline and wall releases, respectively. Solid lines -- numerical results, symbols -- experimental data.}
\end{figure*}
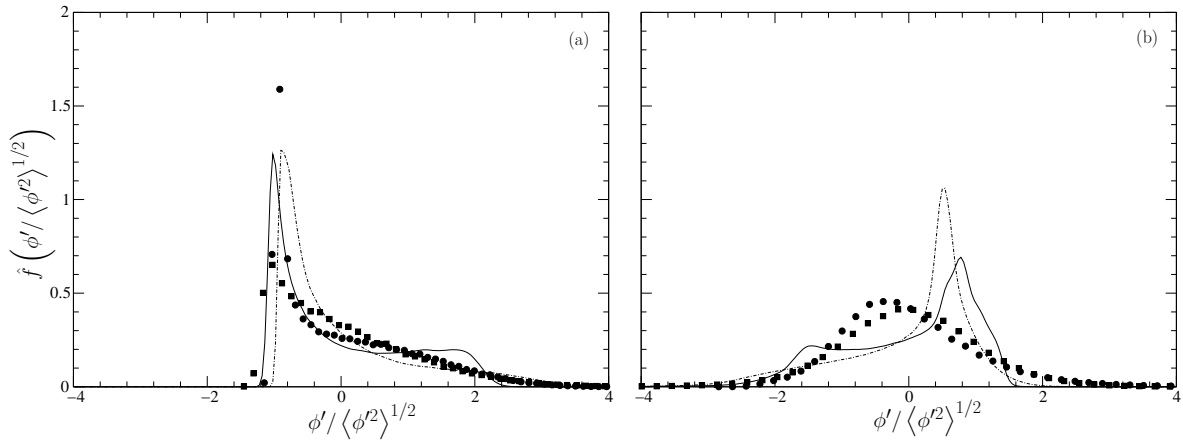
\begin{figure*}
\centering
\resizebox{15.5cm}{!}{\input{pdfs.pstex_t}}
\caption{\label{fig:pdfs}
Probability density functions of scalar concentration fluctuations at selected downstream locations for the (a) centerline and (b) wall-releases at the cross-stream location of their respective sources (i.e.\ \(y/h=1.0\) and \(y/h=0.067\), respectively). Lines -- calculation, symbols -- experimental data at solid line, \(\bullet\), \(x/h=4.0\) and dot-dashed line, \(\blacksquare\), \(x/h=10.8\).}
\end{figure*}
Figure \ref{fig:downstream-evolutions} shows downstream evolutions of the peak of the mean and r.m.s.\ and the width of the mean concentration profiles. In homogeneous isotropic turbulence and homogeneous turbulent shear flow the decay rate of the peak of the mean concentration profiles is reasonably well described by a power law of the form \(\mean{\phi}_\mathrm{peak}\propto x^n\). In the present inhomogeneous flow \citet{Lavertu_05}, based on the experiments, suggest decay exponents of \(n\sim-0.7\) and \(-0.6\) for the wall and centerline sources, respectively. These evolutions are compared to experimental data in \Fige{fig:downstream-evolutions} (a) and (b). Downstream evolutions of the width of the mean concentration profiles \(\sigma_\mathrm{mean}\) are plotted in \Fige{fig:downstream-evolutions} (c) and (d) for the two releases. According to the experimental data, these do not exhibit power-law dependence, as is the case in homogeneous flows. Since the simulations are carried out only on the first half of the measured channel length, the three downstream locations are not sufficient to unambiguously decide whether the simulation data exhibits power-law behavior for the peaks and widths of the mean profiles.

The downstream decay of the peak values of the r.m.s.\ profiles can be well-approximated by a power-law of the form \(\mean{\phi'^2}^{\scriptscriptstyle 1/2}_\mathrm{peak}\propto x^n\), similarly to homogeneous shear flow and isotropic grid-generated turbulence, \Fige{fig:downstream-evolutions} (e) and (f). The experiments suggest \(n=-1\) for both releases.

Probability density functions of scalar concentration fluctuations are depicted in \Fige{fig:pdfs} for both release cases. The cross-stream location of these PDFs are chosen to coincide with that of their respective sources, i.e.\ \(y/h=1.0\) for the centerline release and \(y/h=0.067\) for the wall-release. Two downstream locations are plotted, at the first and at the third location from the sources measured by \citet{Lavertu_05}, at \(x/h=4.0\) and \(x/h=10.8\), respectively. While the PDFs for the centerline release are in reasonable agreement with the experiments, some discrepancies are apparent in the wall-release case. A possible reason behind this disparity is the ad-hoc specification of the mixing timescale in \Eqre{eq:micromixing-timescale}, which is mostly based on theoretical considerations and experimental observations in homogeneous turbulence.

\subsection{Conditional statistics}
\label{sec:res_cond_stat}
The current model solves for the full joint PDF of the turbulent velocity, frequency and scalar concentration. Therefore we can also examine those quantities that require closure assumptions in composition-only PDF methods. This methods are often used in combustion engineering to model complex chemical reactions in a given turbulent flow or in dispersion modeling in the atmospheric boundary layer. In these cases the simplest approach is to assume the shape of the velocity PDF and numerically solve a set of coupled model equations that govern the evolution of the joint PDF of the individual species concentrations in composition space.
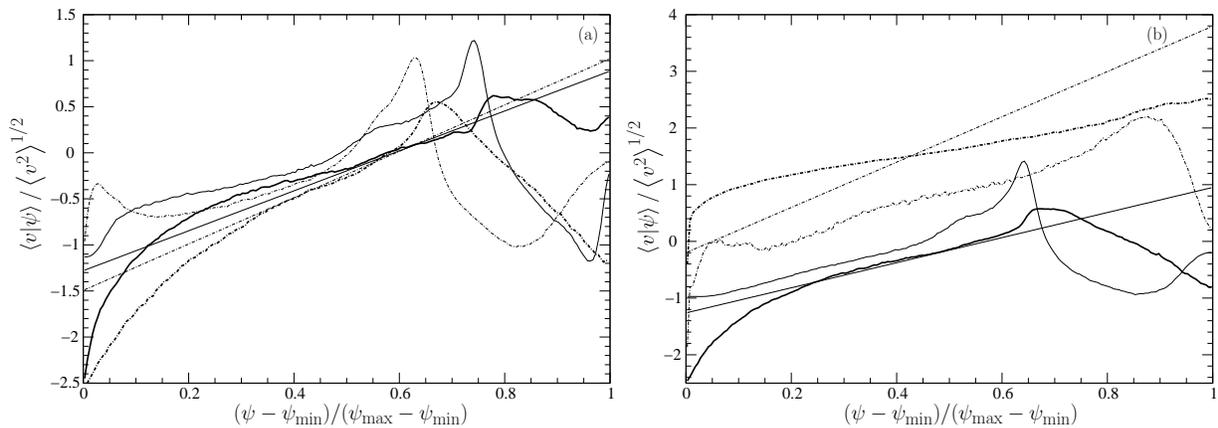
\begin{figure*}
\centering
\resizebox{16cm}{!}{\input{cond_vel.pstex_t}}
\caption{\label{fig:conditioned-velocity}Cross-stream velocity fluctuation conditioned on the scalar concentration for the wall-release (\(y_s/h=0.067\)). Thick lines, IECM model; thin lines, gradient diffusion approximation of \Eqre{eq:gradient-diffusion}; straight sloping lines, linear approximation of \Eqre{eq:linear-conditioned-velocity}. (a) downstream evolution at the height of the source \(y/h=0.067\): solid lines, \(x/h=4.0\); dot-dashed lines, \(x/h=10.8\) and (b) cross-stream evolution at \(x/h=7.4\): solid lines, \(y/h=0.067\); dot-dashed lines, \(y/h=0.67\).}
\end{figure*}
The conservation equation for a single reactive scalar is
\begin{equation}
\frac{\partial\phi}{\partial t} + \bv{U}\cdot\nabla\phi = \Gamma\nabla^2\phi + S(\phi(\bv{x},t)),
\end{equation}
where \(S(\phi)\) is the chemical source term. In high Reynolds number, constant property flow the PDF of a reactive scalar \(f_\phi(\psi;\bv{x},t)\) is governed by \citep{Dopazo_94,Pope_00}
\begin{equation}
\begin{split}
\frac{\partial f_\phi}{\partial t} + \mean{U_i}\frac{\partial f_\phi}{\partial x_i} &= \Gamma\nabla^2f_\phi - \frac{\partial}{\partial x_i}\left(f_\phi\mean{u_i|\psi}\right)\\
&\quad-\frac{\partial^2}{\partial\psi^2}\left(f_\phi\mean{\Gamma\frac{\partial\phi}{\partial x_i}\frac{\partial\phi}{\partial x_i}\bigg|\psi}\right)\\
&\quad-\frac{\partial}{\partial\psi}\left[f_\phi S(\psi)\right],
\end{split}
\end{equation}
or alternatively
\begin{eqnarray}
\lefteqn{\frac{\partial f_\phi}{\partial t} + \frac{\partial}{\partial x_i}\big[f_\phi\left(\mean{U_i}+\mean{u_i|\psi}\right)\big] =}\nonumber\\
&&\qquad\qquad\quad-\frac{\partial}{\partial\psi}\Big\{f_\phi\big[\mean{\Gamma\nabla^2\phi|\psi} + S(\psi)\big]\Big\}.
\end{eqnarray}
An attractive feature of these formulations is that the usually highly nonlinear chemical source term is in closed form. Closure assumptions, however, are necessary for the velocity fluctuations conditional on the scalar concentration \(\mean{u_i|\psi}\) and the conditional scalar dissipation \(\mean{2\Gamma\nabla\phi\cdot\nabla\phi|\psi}\) or the conditional scalar diffusion \(\mean{\Gamma\nabla^2\phi|\psi}\). Since for the current case \(S(\phi)=0\), the marginal scalar PDF \(\Hat{f}(\psi)\) defined in \Eqre{eq:marginal-scalar-pdf} is equal to \(f_\phi\), thus in the remaining text we use them interchangeably.

For the convective term \citet{Dopazo_75} applied the linear approximation
\begin{equation}
\mean{u_i|\psi} = \frac{\mean{u_i\phi'}}{\mean{\phi'^2}}\left(\psi-\mean{\phi}\right),\label{eq:linear-conditioned-velocity}
\end{equation}
to compute the centerline evolution of the temperature PDF in a turbulent axisymmetric heated jet. This linear approximation is exact for joint Gaussian velocity and scalar fluctuations. While many experiments \citep{Bezuglov_74,Golovanov_77,Shcherbina_82,Venkataramani_78,Sreenivasan_78} confirm the linearity of the conditional mean velocity around the local mean conserved scalar, \citet{Kuznetsov_90} observe that most of the experimental data show departure from this linear relationship when \(\left|\psi-\mean{\phi}\right|\) is large. Experimental data from \citet{Sreenivasan_78} and \citet{Bilger_91} also show that in inhomogeneous flows the joint PDF of velocity and scalar is not Gaussian, which makes the above linear approximation dubious in a general case. Nevertheless, this linear model is sometime applied to inhomogeneous scalar fields because of its simplicity.
\begin{figure*}
\centering
\resizebox{16cm}{!}{\input{cdiss.pstex_t}}
\caption{\label{fig:conditional-dissipation}
IECM model predictions for the mean scalar dissipation conditioned on the concentration for (a) the centerline release (\(y_s/h=1.0\)) and (b) the wall-release (\(y_s/h=0.067\)) at different downstream locations: solid line, \(x/h=4.0\); dashed line, \(x/h=7.4\) and dot-dashed line, \(x/h=10.8\). The cross-stream locations are the same as the respective source positions. Note the different scales for the dissipation curves between the different releases. Also shown are the scalar PDFs at the same locations for both releases in (c) and (d), respectively.
}
\end{figure*}
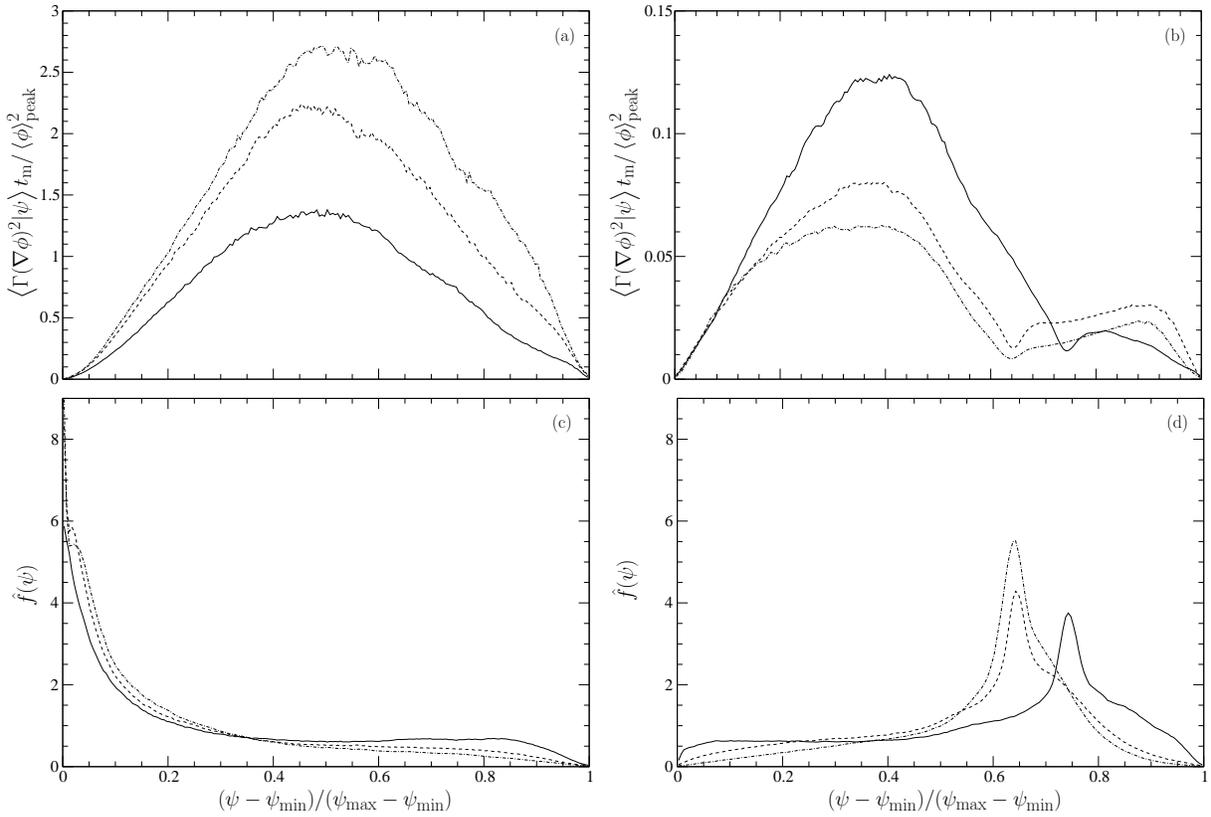

Another commonly employed approximation is to invoke the gradient diffusion hypothesis
\begin{equation}
-f_\phi\mean{u_i|\psi}=\Gamma_T\frac{\partial f_\phi}{\partial x_i},\label{eq:gradient-diffusion}
\end{equation}
where \(\Gamma_T(\bv{x},t)\) is the turbulent diffusivity. In the current case, we specify the turbulent viscosity \(\nu_T\) based on the traditional \(k-\varepsilon\) closure and relate it to \(\Gamma_T\) with the turbulent Prandtl number \(\sigma_T\) as
\begin{equation}
\Gamma_T=\frac{\nu_T}{\sigma_T}=\frac{C_\mu}{\sigma_T}\frac{k^2}{\varepsilon},
\end{equation}
where \(C_\mu=0.09\) is the usual constant in the \(k-\varepsilon\) model and we choose \(\sigma_T=0.8\).

In \Fige{fig:conditioned-velocity} (a) the downstream evolution of the cross-stream velocity fluctuation conditioned on the scalar is depicted for the wall-release case. Both locations are at the height of the source, i.e.\ \(y/h=0.067\). The concentration axis for both locations is scaled between their respective local minimum and maximum concentration values, \(\psi_\mathrm{min}\) and \(\psi_\mathrm{max}\). Note that the model curves show higher negative velocity for low-concentration particles as the distance from the source increases. This is expected, since particles deep inside the plume can have very low concentrations only if they did not come from the source but traveled very fast from above, so that they did not have much time to exchange concentration with the source material. As the plume spreads, only particles with stronger negative velocity can maintain their low concentration values. Likewise, as the center of the plume moves towards the centerline of the channel, high-concentration particles also need to have stronger negative velocities to escape from exchange during their journey from the plume-center to our sensors, which is apparent on the right side of the figure. Obviously, the linear approximation (\ref{eq:linear-conditioned-velocity}) cannot be expected to capture the non-linearity of the model curves, but except for extremely low and high concentrations it performs reasonably well. On the other hand, the gradient diffusion approximation is capable of capturing most features of the IECM model behavior: it successfully reproduces the non-linearity, with some discrepancy at low and high concentrations. It is also apparent that the numerical computation of the derivatives of the PDFs in the gradient diffusion model (\ref{eq:gradient-diffusion}) is most sensitive to sampling errors at the concentration extremes due to lower number of particles falling into the concentration bins there.

The cross-stream evolution of the conditioned velocity fluctuation is shown in \Fige{fig:conditioned-velocity} (b). Both sensors are now at the downstream location \(x/h=7.4\) with increasing distance from the wall at \(y/h=0.067\) and \(0.67\). As the sensor moves towards the channel centerline, the detected low-concentration particles need weaker negative velocity to maintain those low concentrations. The sensor locations relative to the plume centerline can be identified by examining the cross-stream velocity of the high concentration particles. The sensors at \(y/h=0.067\) and \(0.67\) are below and above the plume centerline, respectively, since high-concentration particles at these locations possess negative and positive cross-stream velocities. As is expected, the linear approximation reasonably represents the model behavior for mid-concentrations, while its performance degrades at locations with higher non-Gaussianity, i.e.\ towards the edge of the plume. The performance of the gradient diffusion model is reasonable, except at the concentrations extremes.
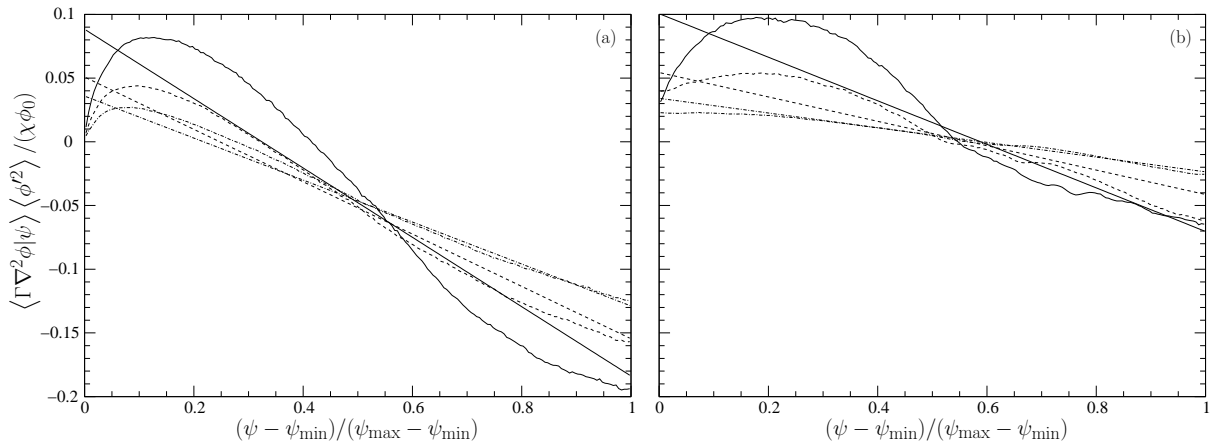
\begin{figure*}
\centering
\resizebox{16cm}{!}{\input{cdiff.pstex_t}}
\caption{\label{fig:conditional-diffusion}
Mean scalar diffusion conditioned on the concentration as predicted by the IECM and IEM models for (a) the centerline release (\(y_s/h=1.0\)) and (b) the wall-release (\(y_s/h=0.067\)) at different downstream locations. The cross-stream locations are the same as the respective source positions. Solid line, \(x/h=4.0\); dashed line, \(x/h=7.4\) and dot-dashed line, \(x/h=10.8\). The straight lines are the linear predictions of the IEM model of \Eqre{eq:IEM-cdiff}.
}
\end{figure*}

For the IECM micromixing model, the mean dissipation conditioned on the scalar concentration can be computed from \citep{Sawford_04b}
\begin{equation}
\mean{2\Gamma\frac{\partial\phi}{\partial x_i}\frac{\partial\phi}{\partial x_i}\bigg|\psi}f_\phi = -\frac{2}{t_\mathrm{m}}\int_0^\psi(\psi'-\Tilde{\phi})f_\phi(\psi')\mathrm{d}\psi',\label{eq:cond-diss}
\end{equation}
where
\begin{equation}
\Tilde{\phi}(\psi)=\int\mean{\phi|\bv{V}}f(\bv{V}|\psi)\mathrm{d}\bv{V}.\label{eq:double-conditioned-scalar}
\end{equation}
The function \(\Tilde{\phi}(\psi)\) in \Eqre{eq:double-conditioned-scalar} can be obtained by taking the average of \(\mean{\phi|\bv{V}}\) over those particles that reside in the bin centered on \(\psi\). The integral in \Eqre{eq:cond-diss}, however, is more problematic. As \citet{Sawford_04b} notes, numerical integration errors that accumulate at extreme concentrations may be amplified when divided by the scalar PDF approaching zero at those locations. Since the integral over all concentrations vanishes, i.e.\ \(\mean{2\Gamma(\nabla\phi)^2|\psi_\mathrm{max}}f_\phi(\psi_\mathrm{max})=0\), for mid-concentrations it can be evaluated either from the left (\(\psi_\mathrm{min}\to\psi\)) or from the right (\(\psi_\mathrm{max}\to\psi\)). Thus the integration errors at the concentration extremes can be significantly decreased by dividing the domain into two parts, integrating the left side from the left and the right side from the right and merging the two results in the division-point. Due to statistical errors, however, the integral over all concentrations may not vanish. In that case, the nonzero value
\begin{equation}
\int_{\psi_\mathrm{min}}^{\psi_\mathrm{max}}(\psi-\Tilde{\phi})f_\phi(\psi)\mathrm{d}\psi
\end{equation}
can be distributed over the sample space by correcting the integrand with the appropriate fraction of this error in each bin.

The conditional mean dissipation for three different downstream locations is depicted in \Fige{fig:conditional-dissipation} for both release cases. As for the conditional velocity, the abscissas here are also scaled between the local \(\psi_\mathrm{min}\) and \(\psi_\mathrm{max}\). The dissipation is normalized by the mixing timescale \(t_\mathrm{m}\) and the square of the mean scalar peak \(\mean{\phi}_\mathrm{peak}^2\) at the corresponding downstream locations. Note that in the case of the wall-release, the dissipation curves are an order of magnitude lower than in the centerline release case. This is mainly a result of the choice of the different micromixing model constants, especially \(C_t\).

In the case of the wall release, the curves exhibit bi-modal shapes at all three downstream locations. This tendency has also been observed by \citet{Kailasnath_93} in the wake of a cylinder and by Sawford in a double-scalar mixing layer \citep{Sawford_06} and, to a lesser extent, also in homogeneous turbulence \citep{Sawford_04b}. \citet{Sardi_98} suggest that in assumed-PDF methods of turbulent combustion a qualitative representation of the conditional dissipation can be obtained in terms of the inverse PDF. To examine this relationship, the corresponding scalar PDFs are also plotted in \Fige{fig:conditional-dissipation} with the same scaling on the concentration axis as the dissipation curves. It is apparent that these results support this reciprocal connection except at the extremes: high values of the PDF correspond to low dissipation (and vice versa). This can be observed for both releases, but it is most visible in the wall-release case, where the mid-concentration minimum between the two maxima of the bi-modal dissipation curves correspond to the peaks in the PDFs.

The IECM model \Eqr{eq:IECM} implies a model for the mean diffusion conditioned on the scalar concentration as
\begin{equation}
\big<\Gamma\nabla^2\phi\big|\bv{V},\psi\big> = -\frac{1}{t_\mathrm{m}}\big(\psi-\mean{\phi|\bv{V}}\big).
\end{equation}
The downstream evolution of the conditional diffusion is depicted in \Fige{fig:conditional-diffusion} for both releases. The concentration axes are scaled as before and the curves are normalized by the scalar variance \(\mean{\phi'^2}\), the concentration at the source, \(\phi_0\) and the mean unconditioned dissipation \(\mean{\chi}=\mean{2\Gamma(\nabla\phi)^2}\), which is computed by integrating \Eqre{eq:cond-diss} over the whole concentration  space. Also shown are the predictions according to the IEM model, which is given by the linear relationship \citep{Sawford_06}
\begin{equation}
\frac{\mean{\Gamma\nabla^2\phi|\psi}\mean{\phi'^2}}{\chi\phi_0} = \frac{1}{2}\left(\frac{\mean{\phi}}{\phi_0}-\psi\right).\label{eq:IEM-cdiff}
\end{equation}
Far downstream as the scalar gets better mixed, the predictions of the IEM and IECM models get closer. This behavior has been observed for other statistics, as well as for other flows such as the double-scalar mixing layer \citep{Sawford_06}. \citet{Kailasnath_93} report experimental data on similar shapes for the conditional diffusion in the turbulent wake of a cylinder.

\section{Conclusions}
\label{sec:conclusion}
Several different stochastic models have been combined to develop a complete PDF-IECM model to compute the joint PDF of turbulent velocity, frequency and scalar concentration in fully developed turbulent channel flow. The flow is represented by a large number of Lagrangian particles and the governing stochastic differential equations have been integrated in time in a Monte-Carlo fashion. The high anisotropy and inhomogeneity at the low-Reynolds-number wall-region have been captured through the elliptic relaxation technique, explicitly modeling the vicinity of the wall down to the viscous sublayer by imposing only the no-slip condition. \citet{Durbin_93} suggested the simple LRR-IP closure of \citet{Launder_75}, originally developed in the Eulerian framework, as a local model used in the elliptic relaxation equation \Eqr{eq:elliptic-relaxation-Lagrangian}. Since then, several more sophisticated local Reynolds stress models have been investigated in conjunction with the elliptic relaxation technique \citep{Whizman_96}. In the PDF framework, the Lagrangian modified IP model of \citet{Pope_94} is based on the LRR-IP closure. We introduced an additional model constant \(C_\xi\) in the definition of the characteristic lengthscale \(L\) \Eqr{eq:L} whose curvature determines the behavior of the relaxation and, ultimately, the overall performance of the model in representing the Reynolds stress anisotropy. This resulted in a correction of the original model overprediction of the wall-normal component \(\mean{v^{\scriptscriptstyle 2}}\) far from the wall, which crucially influences the cross-stream mixing of the transported scalar. However, increasing the constant \(C_\xi\) adversely affects the level of anisotropy that can be represented by the technique. A more accurate treatment of the Reynolds stresses and scalar mixing should be achieved by a more elaborate second moment closure, such as the nonlinear C-L model of \citet{Craft_91} or the Lagrangian version of the SSG model of \citet{Speziale_91} suggested by \citet{Pope_94}.

An unstructured triangular grid is used to compute Eulerian scalar statistics and to track particles throughout the domain. The main purpose of employing unstructured grids has been to prepare the methodology for more complex flow geometries. A similar particle-in-cell approach has been developed by \citet{Muradoglu_99,Muradoglu_01} and by \citet{Zhang_04} for the computation of turbulent reactive flows. These approaches combine the advantages of traditional Eulerian CFD codes with PDF methods in a hybrid manner. Our aim here is to develop a method that is not a hybrid one, so the consistency between the computed fields can be naturally ensured. The emphasis is placed on generality, employing numerical techniques that assume as little as possible about the shape of the numerically computed joint PDF.

We compared the performance of the IEM and the IECM micromixing models in an inhomogeneous flow with strong viscous effects by modeling both the turbulent velocity field and the scalar mixing. The more sophisticated IECM model provides a closer agreement with experimental data in channel flow for the additional computational expense of 30-40\% compared to the IEM model.

Several conditional statistics that often require closure assumptions in PDF models where the velocity field is assumed were extracted and compared to some of their closures. In particular, our conclusions suggest that the scalar-conditioned velocity is well approximated by a linear assumption for mid-concentrations at locations where the velocity PDF is moderately skewed. The gradient diffusion approximation, however, captures most features including the nonlinearity and achieves a closer agreement with the IECM model in slightly more skewed regions of the flow as well. At local concentration extremes and in extremely skewed regions the gradient diffusion approximation markedly departs from the IECM model. The mean scalar dissipation conditioned on the scalar concentration may be well-approximated by the inverse relationship suggested by \citet{Sardi_98} in inhomogeneous flows with significant viscous effects as well, except at the concentration extremes. In computing the conditional scalar diffusion, both the IEM and the IECM models produce similar slopes due to the same scalar dissipation rate attained. 

\begin{acknowledgments}
We wish to thank Thomas Dreeben for the helpful and insightful discussions on the velocity model, Rodney Fox for his interesting comments on the IECM model, Hiroyuki Abe and Hiroshi Kawamura for providing the DNS data on the turbulent velocity field and Laurent Mydlarski, who kindly provided the experimental data on the scalar statistics.
\end{acknowledgments}

\bibliographystyle{physfluids.bst}
\bibliography{jbakosi}

\end{document}

%% file: velocity.pstex_t
\begin{picture}(0,0)%
\includegraphics{velocity.pstex}%
\end{picture}%
\setlength{\unitlength}{4144sp}%
\begingroup\makeatletter\ifx\SetFigFontNFSS\undefined%
\gdef\SetFigFontNFSS#1#2#3#4#5{%
  \reset@font\fontsize{#1}{#2pt}%
  \fontfamily{#3}\fontseries{#4}\fontshape{#5}%
  \selectfont}%
\fi\endgroup%
\begin{picture}(22031,14992)(358,-15461)
%  text 
\put(17371,-15316){\makebox(0,0)[lb]{\smash{{\SetFigFontNFSS{29}{34.8}{\familydefault}{\mddefault}{\updefault}{\color[rgb]{0,0,0}\(y^+\)}%
}}}}
%  text 
\put(6166,-15316){\makebox(0,0)[lb]{\smash{{\SetFigFontNFSS{29}{34.8}{\familydefault}{\mddefault}{\updefault}{\color[rgb]{0,0,0}\(y^+\)}%
}}}}
%  text 
\put(20026,-1456){\makebox(0,0)[lb]{\smash{{\SetFigFontNFSS{20}{24.0}{\familydefault}{\mddefault}{\updefault}{\color[rgb]{0,0,0}\(k\) DNS}%
}}}}
%  text 
\put(20026,-1816){\makebox(0,0)[lb]{\smash{{\SetFigFontNFSS{20}{24.0}{\familydefault}{\mddefault}{\updefault}{\color[rgb]{0,0,0}\(\mean{u^2}\) DNS}%
}}}}
%  text 
\put(20026,-2176){\makebox(0,0)[lb]{\smash{{\SetFigFontNFSS{20}{24.0}{\familydefault}{\mddefault}{\updefault}{\color[rgb]{0,0,0}\(\mean{v^2}\) DNS}%
}}}}
%  text 
\put(20026,-2941){\makebox(0,0)[lb]{\smash{{\SetFigFontNFSS{20}{24.0}{\familydefault}{\mddefault}{\updefault}{\color[rgb]{0,0,0}\(k\) model}%
}}}}
%  text 
\put(20026,-3301){\makebox(0,0)[lb]{\smash{{\SetFigFontNFSS{20}{24.0}{\familydefault}{\mddefault}{\updefault}{\color[rgb]{0,0,0}\(\mean{u^2}\) model}%
}}}}
%  text 
\put(20026,-3661){\makebox(0,0)[lb]{\smash{{\SetFigFontNFSS{20}{24.0}{\familydefault}{\mddefault}{\updefault}{\color[rgb]{0,0,0}\(\mean{v^2}\) model}%
}}}}
%  text 
\put(20026,-4021){\makebox(0,0)[lb]{\smash{{\SetFigFontNFSS{20}{24.0}{\familydefault}{\mddefault}{\updefault}{\color[rgb]{0,0,0}\(\mean{w^2}\) model}%
}}}}
%  text 
\put(20026,-2581){\makebox(0,0)[lb]{\smash{{\SetFigFontNFSS{20}{24.0}{\familydefault}{\mddefault}{\updefault}{\color[rgb]{0,0,0}\(\mean{w^2}\) DNS}%
}}}}
%  text 
\put(10576,-8341){\makebox(0,0)[lb]{\smash{{\SetFigFontNFSS{20}{24.0}{\familydefault}{\mddefault}{\updefault}{\color[rgb]{0,0,0}(c)}%
}}}}
%  text 
\put(21691,-8341){\makebox(0,0)[lb]{\smash{{\SetFigFontNFSS{20}{24.0}{\familydefault}{\mddefault}{\updefault}{\color[rgb]{0,0,0}(d)}%
}}}}
%  text 
\put(21736,-1096){\makebox(0,0)[lb]{\smash{{\SetFigFontNFSS{20}{24.0}{\familydefault}{\mddefault}{\updefault}{\color[rgb]{0,0,0}(b)}%
}}}}
%  text 
\put(10576,-1186){\makebox(0,0)[lb]{\smash{{\SetFigFontNFSS{20}{24.0}{\familydefault}{\mddefault}{\updefault}{\color[rgb]{0,0,0}(a)}%
}}}}
%  text 
\put(11926,-4876){\rotatebox{90.0}{\makebox(0,0)[lb]{\smash{{\SetFigFontNFSS{29}{34.8}{\familydefault}{\mddefault}{\updefault}{\color[rgb]{0,0,0}\(\mean{u_iu_j}/u_\tau^2\)}%
}}}}}
%  text 
\put(766,-4741){\rotatebox{90.0}{\makebox(0,0)[lb]{\smash{{\SetFigFontNFSS{29}{34.8}{\familydefault}{\mddefault}{\updefault}{\color[rgb]{0,0,0}\(\mean{U}/u_\tau\)}%
}}}}}
%  text 
\put(721,-12166){\rotatebox{90.0}{\makebox(0,0)[lb]{\smash{{\SetFigFontNFSS{29}{34.8}{\familydefault}{\mddefault}{\updefault}{\color[rgb]{0,0,0}\(-\mean{uv}/u_\tau^2\)}%
}}}}}
%  text 
\put(11836,-11761){\rotatebox{90.0}{\makebox(0,0)[lb]{\smash{{\SetFigFontNFSS{29}{34.8}{\familydefault}{\mddefault}{\updefault}{\color[rgb]{0,0,0}\(\varepsilon\nu/u_\tau^4\)}%
}}}}}
\end{picture}%

%% file: comparison.pstex_t
\begin{picture}(0,0)%
\includegraphics{comparison.pstex}%
\end{picture}%
\setlength{\unitlength}{3947sp}%
\begingroup\makeatletter\ifx\SetFigFontNFSS\undefined%
\gdef\SetFigFontNFSS#1#2#3#4#5{%
  \reset@font\fontsize{#1}{#2pt}%
  \fontfamily{#3}\fontseries{#4}\fontshape{#5}%
  \selectfont}%
\fi\endgroup%
\begin{picture}(23319,16608)(-673,-15806)
%  text 
\put(4726,-15661){\makebox(0,0)[lb]{\smash{{\SetFigFontNFSS{29}{34.8}{\familydefault}{\mddefault}{\updefault}{\color[rgb]{0,0,0}\(\phi'/\mean{\phi'^2}^{1/2}\)}%
}}}}
%  text 
\put(16651,-15661){\makebox(0,0)[lb]{\smash{{\SetFigFontNFSS{29}{34.8}{\familydefault}{\mddefault}{\updefault}{\color[rgb]{0,0,0}\(\phi'/\mean{\phi'^2}^{1/2}\)}%
}}}}
%  text 
\put(11551,-12661){\rotatebox{90.0}{\makebox(0,0)[lb]{\smash{{\SetFigFontNFSS{29}{34.8}{\familydefault}{\mddefault}{\updefault}{\color[rgb]{0,0,0}\(\Hat{f}\left(\phi'/\mean{\phi'^2}^{1/2}\right)\)}%
}}}}}
%  text 
\put(-299,-12661){\rotatebox{90.0}{\makebox(0,0)[lb]{\smash{{\SetFigFontNFSS{29}{34.8}{\familydefault}{\mddefault}{\updefault}{\color[rgb]{0,0,0}\(\Hat{f}\left(\phi'/\mean{\phi'^2}^{1/2}\right)\)}%
}}}}}
\put(10051, 89){\makebox(0,0)[lb]{\smash{{\SetFigFontNFSS{20}{24.0}{\familydefault}{\mddefault}{\updefault}{\color[rgb]{0,0,0}(a)}%
}}}}
\put(21901,164){\makebox(0,0)[lb]{\smash{{\SetFigFontNFSS{20}{24.0}{\familydefault}{\mddefault}{\updefault}{\color[rgb]{0,0,0}(b)}%
}}}}
\put(21901,-8236){\makebox(0,0)[lb]{\smash{{\SetFigFontNFSS{20}{24.0}{\familydefault}{\mddefault}{\updefault}{\color[rgb]{0,0,0}(d)}%
}}}}
\put(10051,-8236){\makebox(0,0)[lb]{\smash{{\SetFigFontNFSS{20}{24.0}{\familydefault}{\mddefault}{\updefault}{\color[rgb]{0,0,0}(c)}%
}}}}
%  text 
\put(-224,-4036){\rotatebox{90.0}{\makebox(0,0)[lb]{\smash{{\SetFigFontNFSS{29}{34.8}{\familydefault}{\mddefault}{\updefault}{\color[rgb]{0,0,0}\(\mean{\phi}/\mean{\phi}_\mathrm{peak}\)}%
}}}}}
%  text 
\put(5251,-7186){\makebox(0,0)[lb]{\smash{{\SetFigFontNFSS{29}{34.8}{\familydefault}{\mddefault}{\updefault}{\color[rgb]{0,0,0}\(y/h\)}%
}}}}
%  text 
\put(17101,-7186){\makebox(0,0)[lb]{\smash{{\SetFigFontNFSS{29}{34.8}{\familydefault}{\mddefault}{\updefault}{\color[rgb]{0,0,0}\(y/h\)}%
}}}}
%  text 
\put(11551,-3961){\rotatebox{90.0}{\makebox(0,0)[lb]{\smash{{\SetFigFontNFSS{29}{34.8}{\familydefault}{\mddefault}{\updefault}{\color[rgb]{0,0,0}\(\mean{\phi}/\mean{\phi}_\mathrm{peak}\)}%
}}}}}
\end{picture}%

%% file: cross_stream_stats_condensed.pstex_t
\begin{picture}(0,0)%
\includegraphics{cross_stream_stats_condensed.pstex}%
\end{picture}%
\setlength{\unitlength}{4144sp}%
\begingroup\makeatletter\ifx\SetFigFontNFSS\undefined%
\gdef\SetFigFontNFSS#1#2#3#4#5{%
  \reset@font\fontsize{#1}{#2pt}%
  \fontfamily{#3}\fontseries{#4}\fontshape{#5}%
  \selectfont}%
\fi\endgroup%
\begin{picture}(21211,29601)(-631,-28784)
%  text 
\put( 23,-28104){\makebox(0,0)[lb]{\smash{{\SetFigFontNFSS{17}{20.4}{\familydefault}{\mddefault}{\updefault}{\color[rgb]{0,0,0}\(10^0\)}%
}}}}
%  text 
\put(15304,-28609){\makebox(0,0)[lb]{\smash{{\SetFigFontNFSS{29}{34.8}{\familydefault}{\mddefault}{\updefault}{\color[rgb]{0,0,0}\(y/h\)}%
}}}}
%  text 
\put( 37,-26404){\makebox(0,0)[lb]{\smash{{\SetFigFontNFSS{17}{20.4}{\familydefault}{\mddefault}{\updefault}{\color[rgb]{0,0,0}\(10^1\)}%
}}}}
%  text 
\put( 23,-24703){\makebox(0,0)[lb]{\smash{{\SetFigFontNFSS{17}{20.4}{\familydefault}{\mddefault}{\updefault}{\color[rgb]{0,0,0}\(10^2\)}%
}}}}
%  text 
\put( 30,-23003){\makebox(0,0)[lb]{\smash{{\SetFigFontNFSS{17}{20.4}{\familydefault}{\mddefault}{\updefault}{\color[rgb]{0,0,0}\(10^3\)}%
}}}}
%  text 
\put( 23,-21303){\makebox(0,0)[lb]{\smash{{\SetFigFontNFSS{17}{20.4}{\familydefault}{\mddefault}{\updefault}{\color[rgb]{0,0,0}\(10^4\)}%
}}}}
%  text 
\put(-257,-3405){\rotatebox{90.0}{\makebox(0,0)[lb]{\smash{{\SetFigFontNFSS{29}{34.8}{\familydefault}{\mddefault}{\updefault}{\color[rgb]{0,0,0}\(\mean{\phi}/\mean{\phi}_\mathrm{peak}\)}%
}}}}}
%  text 
\put(5014,-28639){\makebox(0,0)[lb]{\smash{{\SetFigFontNFSS{29}{34.8}{\familydefault}{\mddefault}{\updefault}{\color[rgb]{0,0,0}\(y/h\)}%
}}}}
%  text 
\put(-257,-11325){\rotatebox{90.0}{\makebox(0,0)[lb]{\smash{{\SetFigFontNFSS{29}{34.8}{\familydefault}{\mddefault}{\updefault}{\color[rgb]{0,0,0}\(\mean{\phi'^2}^{1/2}/\mean{\phi'^2}^{1/2}_\mathrm{peak}\)}%
}}}}}
%  text 
\put(-167,-25455){\rotatebox{90.0}{\makebox(0,0)[lb]{\smash{{\SetFigFontNFSS{29}{34.8}{\familydefault}{\mddefault}{\updefault}{\color[rgb]{0,0,0}\(\mean{\phi'^4}/\mean{\phi'^2}^2\)}%
}}}}}
%  text 
\put(-167,-18660){\rotatebox{90.0}{\makebox(0,0)[lb]{\smash{{\SetFigFontNFSS{29}{34.8}{\familydefault}{\mddefault}{\updefault}{\color[rgb]{0,0,0}\(\mean{\phi'^3}/\mean{\phi'^2}^{3/2}\)}%
}}}}}
\put(19801,-14416){\makebox(0,0)[lb]{\smash{{\SetFigFontNFSS{20}{24.0}{\familydefault}{\mddefault}{\updefault}{\color[rgb]{0,0,0}(g)}%
}}}}
\put(9541,-14371){\makebox(0,0)[lb]{\smash{{\SetFigFontNFSS{20}{24.0}{\familydefault}{\mddefault}{\updefault}{\color[rgb]{0,0,0}(c)}%
}}}}
\put(9496,-7171){\makebox(0,0)[lb]{\smash{{\SetFigFontNFSS{20}{24.0}{\familydefault}{\mddefault}{\updefault}{\color[rgb]{0,0,0}(b)}%
}}}}
\put(19846,-7171){\makebox(0,0)[lb]{\smash{{\SetFigFontNFSS{20}{24.0}{\familydefault}{\mddefault}{\updefault}{\color[rgb]{0,0,0}(f)}%
}}}}
\put(19801,164){\makebox(0,0)[lb]{\smash{{\SetFigFontNFSS{20}{24.0}{\familydefault}{\mddefault}{\updefault}{\color[rgb]{0,0,0}(e)}%
}}}}
\put(9496,164){\makebox(0,0)[lb]{\smash{{\SetFigFontNFSS{20}{24.0}{\familydefault}{\mddefault}{\updefault}{\color[rgb]{0,0,0}(a)}%
}}}}
\put(19819,-21619){\makebox(0,0)[lb]{\smash{{\SetFigFontNFSS{20}{24.0}{\familydefault}{\mddefault}{\updefault}{\color[rgb]{0,0,0}(h)}%
}}}}
\put(9559,-21664){\makebox(0,0)[lb]{\smash{{\SetFigFontNFSS{20}{24.0}{\familydefault}{\mddefault}{\updefault}{\color[rgb]{0,0,0}(d)}%
}}}}
\end{picture}%

%% file: downstream_evolutions_condensed.pstex_t
\begin{picture}(0,0)%
\includegraphics{downstream_evolutions_condensed.pstex}%
\end{picture}%
\setlength{\unitlength}{4144sp}%
\begingroup\makeatletter\ifx\SetFigFontNFSS\undefined%
\gdef\SetFigFontNFSS#1#2#3#4#5{%
  \reset@font\fontsize{#1}{#2pt}%
  \fontfamily{#3}\fontseries{#4}\fontshape{#5}%
  \selectfont}%
\fi\endgroup%
\begin{picture}(21097,22660)(-508,-21892)
%  text 
\put(420,-21280){\makebox(0,0)[lb]{\smash{{\SetFigFontNFSS{17}{20.4}{\familydefault}{\mddefault}{\updefault}{\color[rgb]{0,0,0}\(10^0\)}%
}}}}
%  text 
\put(10680,-21280){\makebox(0,0)[lb]{\smash{{\SetFigFontNFSS{17}{20.4}{\familydefault}{\mddefault}{\updefault}{\color[rgb]{0,0,0}\(10^0\)}%
}}}}
%  text 
\put(5285,-21280){\makebox(0,0)[lb]{\smash{{\SetFigFontNFSS{17}{20.4}{\familydefault}{\mddefault}{\updefault}{\color[rgb]{0,0,0}\(10^1\)}%
}}}}
%  text 
\put(15545,-21280){\makebox(0,0)[lb]{\smash{{\SetFigFontNFSS{17}{20.4}{\familydefault}{\mddefault}{\updefault}{\color[rgb]{0,0,0}\(10^1\)}%
}}}}
%  text 
\put(10135,-21280){\makebox(0,0)[lb]{\smash{{\SetFigFontNFSS{17}{20.4}{\familydefault}{\mddefault}{\updefault}{\color[rgb]{0,0,0}\(10^2\)}%
}}}}
%  text 
\put(20395,-21280){\makebox(0,0)[lb]{\smash{{\SetFigFontNFSS{17}{20.4}{\familydefault}{\mddefault}{\updefault}{\color[rgb]{0,0,0}\(10^2\)}%
}}}}
%  text 
\put(15706,-21706){\makebox(0,0)[lb]{\smash{{\SetFigFontNFSS{29}{34.8}{\familydefault}{\mddefault}{\updefault}{\color[rgb]{0,0,0}\(x/h\)}%
}}}}
%  text 
\put(9541,-14686){\makebox(0,0)[lb]{\smash{{\SetFigFontNFSS{20}{24.0}{\familydefault}{\mddefault}{\updefault}{\color[rgb]{0,0,0}(e)}%
}}}}
%  text 
\put(19846,-14731){\makebox(0,0)[lb]{\smash{{\SetFigFontNFSS{20}{24.0}{\familydefault}{\mddefault}{\updefault}{\color[rgb]{0,0,0}(f)}%
}}}}
%  text 
\put(9586, 74){\makebox(0,0)[lb]{\smash{{\SetFigFontNFSS{20}{24.0}{\familydefault}{\mddefault}{\updefault}{\color[rgb]{0,0,0}(a)}%
}}}}
%  text 
\put(19801, 74){\makebox(0,0)[lb]{\smash{{\SetFigFontNFSS{20}{24.0}{\familydefault}{\mddefault}{\updefault}{\color[rgb]{0,0,0}(b)}%
}}}}
%  text 
\put(19846,-7261){\makebox(0,0)[lb]{\smash{{\SetFigFontNFSS{20}{24.0}{\familydefault}{\mddefault}{\updefault}{\color[rgb]{0,0,0}(d)}%
}}}}
%  text 
\put(9541,-7261){\makebox(0,0)[lb]{\smash{{\SetFigFontNFSS{20}{24.0}{\familydefault}{\mddefault}{\updefault}{\color[rgb]{0,0,0}(c)}%
}}}}
%  text 
\put( 45,-5041){\makebox(0,0)[lb]{\smash{{\SetFigFontNFSS{17}{20.4}{\familydefault}{\mddefault}{\updefault}{\color[rgb]{0,0,0}\(10^{-1}\)}%
}}}}
%  text 
\put( 45,-12620){\makebox(0,0)[lb]{\smash{{\SetFigFontNFSS{17}{20.4}{\familydefault}{\mddefault}{\updefault}{\color[rgb]{0,0,0}\(10^{-1}\)}%
}}}}
%  text 
\put(5311,-21751){\makebox(0,0)[lb]{\smash{{\SetFigFontNFSS{29}{34.8}{\familydefault}{\mddefault}{\updefault}{\color[rgb]{0,0,0}\(x/h\)}%
}}}}
%  text 
\put( 95,525){\makebox(0,0)[lb]{\smash{{\SetFigFontNFSS{17}{20.4}{\familydefault}{\mddefault}{\updefault}{\color[rgb]{0,0,0}\(10^0\)}%
}}}}
%  text 
\put( 95,-8154){\makebox(0,0)[lb]{\smash{{\SetFigFontNFSS{17}{20.4}{\familydefault}{\mddefault}{\updefault}{\color[rgb]{0,0,0}\(10^0\)}%
}}}}
%  text 
\put( 75,-21036){\makebox(0,0)[lb]{\smash{{\SetFigFontNFSS{17}{20.4}{\familydefault}{\mddefault}{\updefault}{\color[rgb]{0,0,0}\(10^{-2}\)}%
}}}}
%  text 
\put( 90,-17635){\makebox(0,0)[lb]{\smash{{\SetFigFontNFSS{17}{20.4}{\familydefault}{\mddefault}{\updefault}{\color[rgb]{0,0,0}\(10^{-1}\)}%
}}}}
%  text 
\put(140,-14235){\makebox(0,0)[lb]{\smash{{\SetFigFontNFSS{17}{20.4}{\familydefault}{\mddefault}{\updefault}{\color[rgb]{0,0,0}\(10^0\)}%
}}}}
%  text 
\put(-44,-10681){\rotatebox{90.0}{\makebox(0,0)[lb]{\smash{{\SetFigFontNFSS{29}{34.8}{\familydefault}{\mddefault}{\updefault}{\color[rgb]{0,0,0}\(\sigma_\mathrm{mean}/h\)}%
}}}}}
%  text 
\put(-89,-3571){\rotatebox{90.0}{\makebox(0,0)[lb]{\smash{{\SetFigFontNFSS{29}{34.8}{\familydefault}{\mddefault}{\updefault}{\color[rgb]{0,0,0}\(\mean{\phi}_\mathrm{peak}\)}%
}}}}}
%  text 
\put(-134,-18241){\rotatebox{90.0}{\makebox(0,0)[lb]{\smash{{\SetFigFontNFSS{29}{34.8}{\familydefault}{\mddefault}{\updefault}{\color[rgb]{0,0,0}\(\mean{\phi'^2}^{1/2}_\mathrm{peak}\)}%
}}}}}
\end{picture}%

%% file: pdfs.pstex_t
\begin{picture}(0,0)%
\includegraphics{pdfs.pstex}%
\end{picture}%
\setlength{\unitlength}{4144sp}%
\begingroup\makeatletter\ifx\SetFigFontNFSS\undefined%
\gdef\SetFigFontNFSS#1#2#3#4#5{%
  \reset@font\fontsize{#1}{#2pt}%
  \fontfamily{#3}\fontseries{#4}\fontshape{#5}%
  \selectfont}%
\fi\endgroup%
\begin{picture}(21160,7863)(-598,-7046)
%  text 
\put(4681,-6901){\makebox(0,0)[lb]{\smash{{\SetFigFontNFSS{29}{34.8}{\familydefault}{\mddefault}{\updefault}{\color[rgb]{0,0,0}\(\phi'/\mean{\phi'^2}^{1/2}\)}%
}}}}
%  text 
\put(9406, 74){\makebox(0,0)[lb]{\smash{{\SetFigFontNFSS{20}{24.0}{\familydefault}{\mddefault}{\updefault}{\color[rgb]{0,0,0}(a)}%
}}}}
%  text 
\put(19711,119){\makebox(0,0)[lb]{\smash{{\SetFigFontNFSS{20}{24.0}{\familydefault}{\mddefault}{\updefault}{\color[rgb]{0,0,0}(b)}%
}}}}
%  text 
\put(14941,-6856){\makebox(0,0)[lb]{\smash{{\SetFigFontNFSS{29}{34.8}{\familydefault}{\mddefault}{\updefault}{\color[rgb]{0,0,0}\(\phi'/\mean{\phi'^2}^{1/2}\)}%
}}}}
%  text 
\put(-224,-4201){\rotatebox{90.0}{\makebox(0,0)[lb]{\smash{{\SetFigFontNFSS{29}{34.8}{\familydefault}{\mddefault}{\updefault}{\color[rgb]{0,0,0}\(\Hat{f}\left(\phi'/\mean{\phi'^2}^{1/2}\right)\)}%
}}}}}
\end{picture}%

%% file: cond_vel.pstex_t
\begin{picture}(0,0)%
\includegraphics{cond_vel.pstex}%
\end{picture}%
\setlength{\unitlength}{4144sp}%
\begingroup\makeatletter\ifx\SetFigFontNFSS\undefined%
\gdef\SetFigFontNFSS#1#2#3#4#5{%
  \reset@font\fontsize{#1}{#2pt}%
  \fontfamily{#3}\fontseries{#4}\fontshape{#5}%
  \selectfont}%
\fi\endgroup%
\begin{picture}(22194,7790)(-676,-6956)
\put(9676,209){\makebox(0,0)[lb]{\smash{{\SetFigFontNFSS{20}{24.0}{\familydefault}{\mddefault}{\updefault}{\color[rgb]{0,0,0}(a)}%
}}}}
\put(3331,-6811){\makebox(0,0)[lb]{\smash{{\SetFigFontNFSS{29}{34.8}{\familydefault}{\mddefault}{\updefault}{\color[rgb]{0,0,0}\((\psi-\psi_\mathrm{min})/(\psi_\mathrm{max}-\psi_\mathrm{min})\)}%
}}}}
\put(14581,-6811){\makebox(0,0)[lb]{\smash{{\SetFigFontNFSS{29}{34.8}{\familydefault}{\mddefault}{\updefault}{\color[rgb]{0,0,0}\((\psi-\psi_\mathrm{min})/(\psi_\mathrm{max}-\psi_\mathrm{min})\)}%
}}}}
\put(-314,-3706){\rotatebox{90.0}{\makebox(0,0)[lb]{\smash{{\SetFigFontNFSS{29}{34.8}{\familydefault}{\mddefault}{\updefault}{\color[rgb]{0,0,0}\(\mean{v|\psi}/\mean{v^2}^{1/2}\)}%
}}}}}
\put(11071,-3751){\rotatebox{90.0}{\makebox(0,0)[lb]{\smash{{\SetFigFontNFSS{29}{34.8}{\familydefault}{\mddefault}{\updefault}{\color[rgb]{0,0,0}\(\mean{v|\psi}/\mean{v^2}^{1/2}\)}%
}}}}}
\put(20161,209){\makebox(0,0)[lb]{\smash{{\SetFigFontNFSS{20}{24.0}{\familydefault}{\mddefault}{\updefault}{\color[rgb]{0,0,0}(b)}%
}}}}
\end{picture}%

%% file: cdiss.pstex_t
\begin{picture}(0,0)%
\includegraphics{cdiss.pstex}%
\end{picture}%
\setlength{\unitlength}{4144sp}%
\begingroup\makeatletter\ifx\SetFigFontNFSS\undefined%
\gdef\SetFigFontNFSS#1#2#3#4#5{%
  \reset@font\fontsize{#1}{#2pt}%
  \fontfamily{#3}\fontseries{#4}\fontshape{#5}%
  \selectfont}%
\fi\endgroup%
\begin{picture}(22206,14990)(-598,-14156)
\put(3286,-14011){\makebox(0,0)[lb]{\smash{{\SetFigFontNFSS{29}{34.8}{\familydefault}{\mddefault}{\updefault}{\color[rgb]{0,0,0}\((\psi-\psi_\mathrm{min})/(\psi_\mathrm{max}-\psi_\mathrm{min})\)}%
}}}}
\put(14536,-14011){\makebox(0,0)[lb]{\smash{{\SetFigFontNFSS{29}{34.8}{\familydefault}{\mddefault}{\updefault}{\color[rgb]{0,0,0}\((\psi-\psi_\mathrm{min})/(\psi_\mathrm{max}-\psi_\mathrm{min})\)}%
}}}}
\put(-224,-4651){\rotatebox{90.0}{\makebox(0,0)[lb]{\smash{{\SetFigFontNFSS{29}{34.8}{\familydefault}{\mddefault}{\updefault}{\color[rgb]{0,0,0}\(\mean{\Gamma(\nabla\phi)^2|\psi}t_\mathrm{m}/\mean{\phi}_\mathrm{peak}^2\)}%
}}}}}
\put(10891,-4651){\rotatebox{90.0}{\makebox(0,0)[lb]{\smash{{\SetFigFontNFSS{29}{34.8}{\familydefault}{\mddefault}{\updefault}{\color[rgb]{0,0,0}\(\mean{\Gamma(\nabla\phi)^2|\psi}t_\mathrm{m}/\mean{\phi}_\mathrm{peak}^2\)}%
}}}}}
\put(9496,119){\makebox(0,0)[lb]{\smash{{\SetFigFontNFSS{20}{24.0}{\familydefault}{\mddefault}{\updefault}{\color[rgb]{0,0,0}(a)}%
}}}}
\put(20746,119){\makebox(0,0)[lb]{\smash{{\SetFigFontNFSS{20}{24.0}{\familydefault}{\mddefault}{\updefault}{\color[rgb]{0,0,0}(b)}%
}}}}
\put(-134,-10321){\rotatebox{90.0}{\makebox(0,0)[lb]{\smash{{\SetFigFontNFSS{29}{34.8}{\familydefault}{\mddefault}{\updefault}{\color[rgb]{0,0,0}\(\Hat{f}(\psi)\)}%
}}}}}
\put(10936,-10231){\rotatebox{90.0}{\makebox(0,0)[lb]{\smash{{\SetFigFontNFSS{29}{34.8}{\familydefault}{\mddefault}{\updefault}{\color[rgb]{0,0,0}\(\Hat{f}(\psi)\)}%
}}}}}
\put(9451,-7036){\makebox(0,0)[lb]{\smash{{\SetFigFontNFSS{20}{24.0}{\familydefault}{\mddefault}{\updefault}{\color[rgb]{0,0,0}(c)}%
}}}}
\put(20791,-7036){\makebox(0,0)[lb]{\smash{{\SetFigFontNFSS{20}{24.0}{\familydefault}{\mddefault}{\updefault}{\color[rgb]{0,0,0}(d)}%
}}}}
\end{picture}%

%% file: cdiff.pstex_t
\begin{picture}(0,0)%
\includegraphics{cdiff.pstex}%
\end{picture}%
\setlength{\unitlength}{4144sp}%
\begingroup\makeatletter\ifx\SetFigFontNFSS\undefined%
\gdef\SetFigFontNFSS#1#2#3#4#5{%
  \reset@font\fontsize{#1}{#2pt}%
  \fontfamily{#3}\fontseries{#4}\fontshape{#5}%
  \selectfont}%
\fi\endgroup%
\begin{picture}(21396,7775)(-643,-6956)
\put(3376,-6811){\makebox(0,0)[lb]{\smash{{\SetFigFontNFSS{29}{34.8}{\familydefault}{\mddefault}{\updefault}{\color[rgb]{0,0,0}\((\psi-\psi_\mathrm{min})/(\psi_\mathrm{max}-\psi_\mathrm{min})\)}%
}}}}
\put(13861,-6811){\makebox(0,0)[lb]{\smash{{\SetFigFontNFSS{29}{34.8}{\familydefault}{\mddefault}{\updefault}{\color[rgb]{0,0,0}\((\psi-\psi_\mathrm{min})/(\psi_\mathrm{max}-\psi_\mathrm{min})\)}%
}}}}
\put(-269,-4606){\rotatebox{90.0}{\makebox(0,0)[lb]{\smash{{\SetFigFontNFSS{29}{34.8}{\familydefault}{\mddefault}{\updefault}{\color[rgb]{0,0,0}\(\mean{\Gamma\nabla^2\phi|\psi}\mean{\phi'^2}/(\chi\phi_0)\)}%
}}}}}
\put(9766,164){\makebox(0,0)[lb]{\smash{{\SetFigFontNFSS{20}{24.0}{\familydefault}{\mddefault}{\updefault}{\color[rgb]{0,0,0}(a)}%
}}}}
\put(19981,164){\makebox(0,0)[lb]{\smash{{\SetFigFontNFSS{20}{24.0}{\familydefault}{\mddefault}{\updefault}{\color[rgb]{0,0,0}(b)}%
}}}}
\end{picture}%